\documentclass{aa}

\usepackage{amsmath}
\usepackage{txfonts}
\usepackage{amssymb}
\usepackage{esint}
\usepackage{natbib}
\usepackage{graphicx}
\usepackage{bm}
\usepackage[percent]{overpic}
\usepackage{multirow}
\usepackage[colorlinks=true, linkcolor=blue, citecolor=blue, filecolor=blue, urlcolor=blue]{hyperref}

\newcommand{\authornote}[1]{$^{#1}$}

\begin{document}
\title{Testing the cosmological principle with quasars}
\author{Shuangnan Chen\inst{1,2}
	\and
	Xiaofeng Yang\authornote{*}\inst{2}\thanks{E-mail: xfyang@henu.edu.cn}
	\and
	Yunliang Ren\inst{2,1}
	\and
	Xuwei Zhang\inst{3}
	\and
	Yangjun Shi\inst{2,4}
        \and
        Cheng Cheng\inst{3}
        \and
        Xiaolong He\inst{1}
        \and
        Sufen Guo\authornote{*}\inst{1}\thanks{E-mail: guosufen@xju.edu.cn}
    }
\authorrunning{Chen et al.}
\institute{School of Physics and Technology, Xinjiang University, Urumqi, 830046, China
        \and
        School of Physics and Electronics, Henan University, Kaifeng, 475004, China
	\and
	Xinjiang Astronomical Observatory, Chinese Academy of Sciences, Urumqi, 830011, China
	\and 
	School of Physics and Astronomy, China West Normal University, Sichuan, 637002, China
}

\date{}

\abstract{
The cosmological principle posits that the universe is homogeneous and isotropic on the large scales. In history, the cosmological principle was confirmed by various cosmological observations from CMB to large scale structure. However, several new challenges to the cosmological principle were reported in recent years, particularly in radio observations from overdispersed radio source counts to quasars. Here, we firstly present studies on the peculiar velocity of large-scale anisotropy by measuring the dipole signal from the DESI DR1 catalogue with a sample of 1,176,570 quasars (0.8 < z < 3.0). Our analysis reveals the peculiar velocity of $|v| = 443.8 \pm 204.1$ km/s towards $(l, b) = (107.4^\circ \pm 86.8^\circ, 28.4^\circ \pm 45.2^\circ)$ in Galactic coordinates.The motion direction deviates from the CMB dipole (264.02$^\circ$, 48.253$^\circ$). The inferred velocity is consistent at the 1.56 $\sigma$ level with the value of 370 km/s from a purely kinematic interpretation of the CMB dipole. Based on the motion direction component analysis, we have not found any significant deviation from cosmological principle in current released quasars data. }

\keywords{cosmology principle -- quasars -- peculiar velocity -- large scale structure -- redshift dipole}
\maketitle

\section{Introduction}
\label{sec:intro}

The $\Lambda$ cold dark matter ($\Lambda$CDM) model, validated by decades of observational data, serves as the standard framework of modern cosmology. This model successfully reconciles the observed accelerated expansion of the Universe with the theory of general relativity, resting on two foundational pillars: the theory of gravity itself and the cosmological principle (CP).

The CP, a core postulate of the concordance $\Lambda$CDM model, posits that the Universe is statistically homogeneous and isotropic on scales larger than $\sim 100$ Mpc. This assumption forms the basis of the standard Friedmann-Lemaître-Robertson-Walker (FLRW) cosmology.

Direct support for large-scale isotropy comes from the high degree of uniformity observed in the cosmic microwave background (CMB), which appears remarkably isotropic on small angular scales. Complementary evidence is provided by a variety of independent cosmological tracers, including large-scale galaxy surveys \citep{Gibelyou_2012, Sarkar_2018, Franco_2023}, the X-ray background \citep{Wu_1999}, the spatial distribution of galaxy clusters \citep{Bengaly_2016}, radio galaxies \citep{2002Natur.416..150B}, and Type Ia supernovae \citep{Gupta_2010, sorrenti2023dipolepantheonsh0esdata}. Collectively, these observations robustly corroborate the statistical isotropy of the Universe on the largest observable scales.

Amidst this backdrop of large-scale isotropy, the CMB itself exhibits a hierarchy of anisotropies. It contains small-scale fluctuations at the level of approximately 1 part in $10^5$, and a far more prominent dipole anisotropy of amplitude $\sim 1$ part in $10^3$ as measured in the heliocentric frame. Given its distinctive amplitude and dipolar pattern, this dominant signal is almost universally attributed to a kinematic effect—the peculiar motion of the Solar System relative to the CMB rest frame. The latest measurements infer a velocity of $v = 369.825 \pm 0.11$ km/s towards the Galactic coordinates $(l, b) = (264.021^\circ, 48.253^\circ)$ \citep{Hinshaw_2009, 2020}.

A critical test of the CP involves verifying the purely kinematic origin of the CMB dipole. This is achieved by measuring a corresponding dipole anisotropy in the distribution of distant matter. If the CP holds, the peculiar velocity inferred from the matter rest frame must be consistent with that derived from the CMB. The Ellis–Baldwin test provides a direct implementation of this idea \citep{10.1093/mnras/206.2.377}. It posits that if the CMB dipole is indeed due to our motion through a homogeneous and isotropic universe, then a statistically congruent dipole must be present in the number counts of distant sources, such as radio galaxies. The velocity inferred from these counts should agree with the CMB-derived value. Consequently, under the CP, the large-scale distribution of matter should appear isotropic when viewed from the CMB rest frame. Any significant deviation would necessitate a physical explanation and could potentially challenge the validity of the CP itself.

However, this expectation has been challenged by several findings over the past two decades. \citet{Singal_2011} reported a significant discrepancy, finding the dipole amplitude from discrete radio sources to be approximately four times larger than that of the CMB, albeit with a possibly consistent direction. A congruent result was obtained by \citet{Rubart_2013} using the NVSS and WENSS surveys, who confirmed a dipole direction aligned with the CMB but an amplitude larger by a factor of approximately three. Extending the analysis to Type Ia supernovae, \citet{Singal_2022} similarly found a dipole amplitude about four times greater than the CMB value. In a distinct finding from the Pantheon+SH0ES supernova data, \citet{sorrenti2023dipolepantheonsh0esdata} reported a dipole amplitude consistent with the CMB but a direction differing with a statistical significance exceeding $3\sigma$.Earlier, using the Union2.1 supernovae data, \citet{Yang_2013} also found a preferred dipole direction via a dipole-fitting approach.

\citet{Secrest_2021} used quasars (QSOs) from the CatWISE2020 catalogue and found a dipole direction offset by $\sim 27^\circ$ from the CMB dipole, but with an amplitude over twice as large as expected, rejecting the kinematic interpretation at the $4.9\sigma$ level. Separately, \citet{Dam_2023} also reported a significant tension through a Bayesian analysis of QSO data, measuring a dipole amplitude approximately 2.7 times larger than the kinematic expectation with a statistical significance of $5.7\sigma$. Most recently, an analysis combining NVSS, RACS-low, and LoTSS-DR2 radio surveys reported a dipole amplitude $3.67 \pm 0.49$ times larger than the kinematic expectation, a $5.4\sigma$ discrepancy \citep{B_hme_2025}. \citet{Secrest_2022} further analysed radio galaxies and QSOs; they suggest that if the CMB dipole is assumed to be fully kinematic, the reference frames of these distant sources may exhibit an intrinsic anisotropy. 

Other studies also hint at potential violations of statistical isotropy. For instance, isotropy tests performed on the SDSS Luminous Red Galaxy sample indicate a possible breakdown on scales as large as ${\sim}300 \, h^{-1}$ Mpc \citep{10.1093/mnras/stx988}. In the CMB itself, a persistent hemispherical power asymmetry remains significant at an approximate the $3\sigma$ level in the latest data \citep{Eriksen_2004, refId0, refId01}. Notably, several of these anomalous signals, including the dipole in radio polarization and the alignment of the CMB low-order multipoles, indicate a preferred direction closely aligned with the CMB dipole \citep{RALSTON_2004}. Furthermore, the discovery of enormous cosmic structures—such as large QSO groups \citep{10.1093/mnras/sts497}, correlated QSO orientations \citep{ 10.1093/mnras/stac269}, and giant cosmic voids \citep{Haslbauer_2020, Keenan_2013}—challenges the assumption of homogeneity at the largest scales.

The cumulative evidence from these diverse studies, which utilize different tracers and methodologies and sometimes yield conflicting results, calls into question the foundational assumption of large-scale isotropy and homogeneity inherent in the CP. In this context, we employed an independent cosmological probe to estimate the Solar System's peculiar motion by analysing the dipole anisotropy in the sky distribution of QSO redshifts. The redshift dipole offers a direct measurement of the peculiar velocity relative to the matter reference frame traced by the QSOs. This method circumvents the need to account for local over- or under-densities (e.g - a local void or specific clustering) and other large-scale structures (LSSs) known to potentially bias traditional number-count dipole measurements \citep{2024MNRAS.531.4545O, Rameez_2018, Tiwari_2016, Rubart_2014}.

This study aims to employ the high-precision, wide-coverage QSO data from the Dark Energy Spectroscopic Instrument (DESI) Data Release 1 (DR1) to perform a tomographic estimation of the dipole modulation across redshift bins, following the methodology of \citet{da_Silveira_Ferreira_2024}. We used the DESI DR1 QSO sample to infer the Solar System's peculiar velocity via the aberration effect, thereby performing a direct test of the CP through the Ellis–Baldwin test. This test confronts the kinematic dipole measured in the matter frame (traced by distant QSOs) with that of the cosmic rest frame (defined by the CMB).

This paper is structured as follows. Section~\ref{sec:data} details the DESI and eBOSS QSO catalogues and their sky footprints used in the analysis. Section~\ref{sec:method} presents the formalism for the redshift dipole, outlines the least-squares estimator used to measure the dipole amplitude, justifies the separate analysis of the northern Galactic caps (NGC) and the southern Galactic caps (SGC) to account for systematic depth variations, and introduces Fisher's method for evaluating the overall statistical significance. Finally, Section~\ref{sec:concls} presents the main results and discusses their implications, concluding that the QSO dipole inferred from DESI DR1 is consistent with the kinematic interpretation of the CMB dipole, thereby finding no compelling evidence against the CP.

\section{Data}
\label{sec:data}
We aim to test the CP using LSS tracers from recent spectroscopic surveys. To ensure robust results, we selected catalogues based on data quality, minimal known systematics, substantial sky coverage, and the ability to perform survey comparisons. Specifically, we utilized the final QSO clustering catalogues from the SDSS Extended Baryon Oscillation Spectroscopic Survey (eBOSS) and the early data release of the Dark Energy Spectroscopic Instrument (DESI). These catalogues provide observational weights, redshifts, and angular positions, and their derived cosmological parameters are consistent with the $\Lambda$CDM framework.

We analysed three distinct QSO samples:

(\romannumeral1)
The eBOSS DR14 QSO Catalogue \citep{Abolfathi_2018}: This catalogue contains 194,801 objects in the redshift range $0.04 < z < 6.95$. After removing extreme outliers, 90\% of the data lie within $0.42 < z < 3.28$, leaving 192,853 objects. To mitigate potential boundary effects, we further restricted our analysis to the well-populated range $0.4 < z < 2.8$, resulting in a final sample of 187,954 QSOs. The spectra cover the wavelength region 3610–10140 $Å$ at a spectral resolution in the range $1300 < R < 2500$ \citep{Abolfathi_2018}.

(\romannumeral2)
The eBOSS DR16 QSO Catalogue \citep{10.1093/mnras/staa2416}: This catalogue contains 343,708 QSOs with a tight and uniform redshift distribution in the range $0.8 < z < 2.2$, selected from the final SDSS-IV/eBOSS quasar sample \citep{2020ApJS..250....8L}. We used the entire specified range without further cuts. The spectra cover 3600–10400 $Å$ with a spectral resolution of $R \sim 2000$ \citep{2020ApJS..250....8L}. At the largest QSO sample at the time, eBOSS DR16 provided important data for cosmological analysis.

(\romannumeral3)
The DESI DR1 QSO Catalogue \citep{desicollaboration2025datarelease1dark}: The Dark Energy Spectroscopic Instrument (DESI) is a Stage IV spectroscopic survey designed to obtain tens of millions of galaxy and QSO redshifts to constrain dark energy and cosmological models \citep{Adame_2025}. Its first data release (DR1) encompasses the first year of main survey observations. The public large-scale structure catalogue for QSOs contains 1,223,170 objects with confident redshift measurements in the range $0.8 < z < 3.5$. The spectra cover the wavelength region 3600–9824 $Å$ at a spectral resolution in the range $2000 < R < 5200$ \citep{desicollaboration2025datarelease1dark}.

\begin{figure*}[!htbp]
    \centering
        \includegraphics[width=6cm]{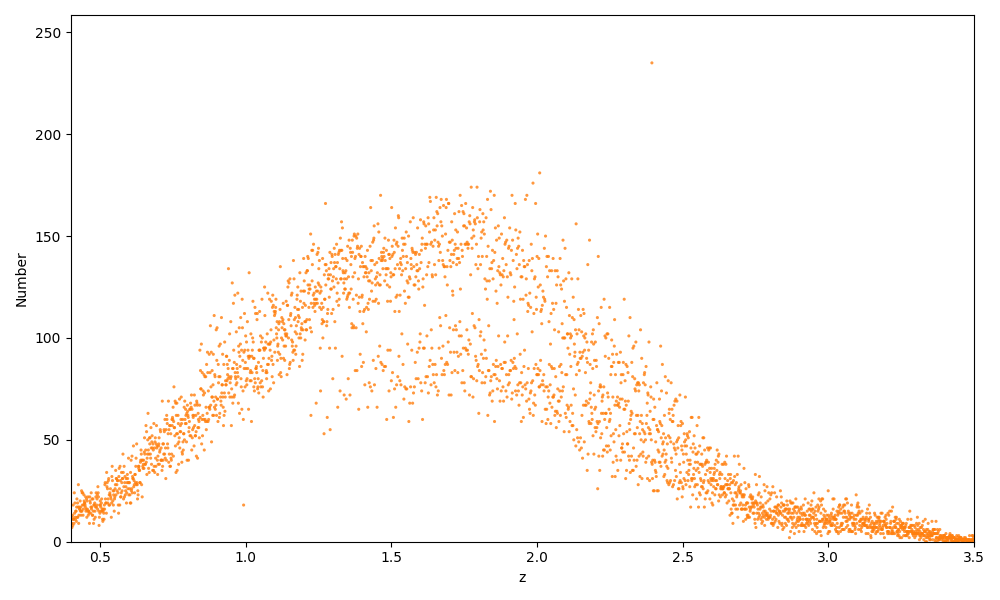}
        \includegraphics[width=6cm]{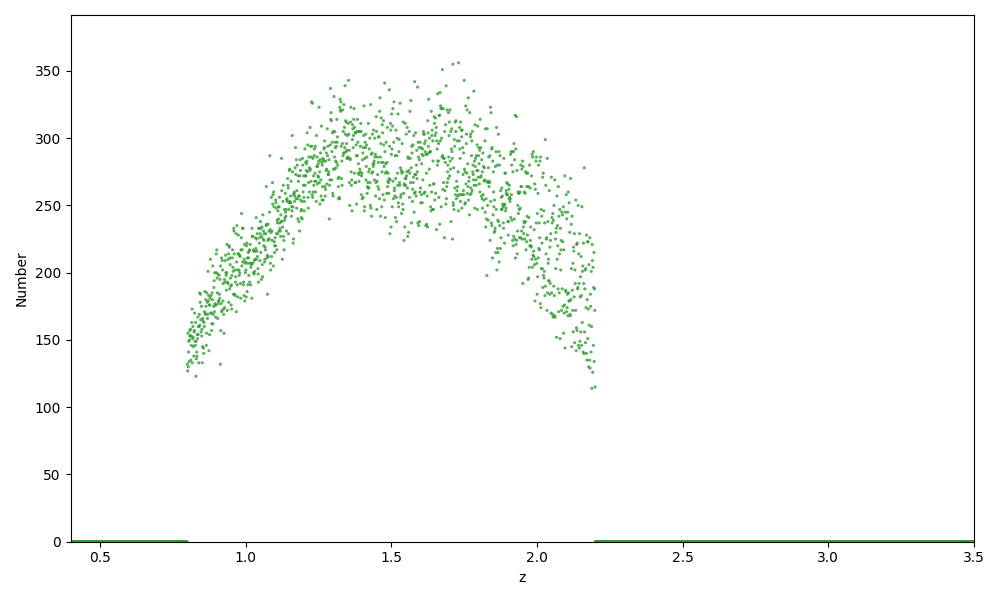}
        \includegraphics[width=6cm]{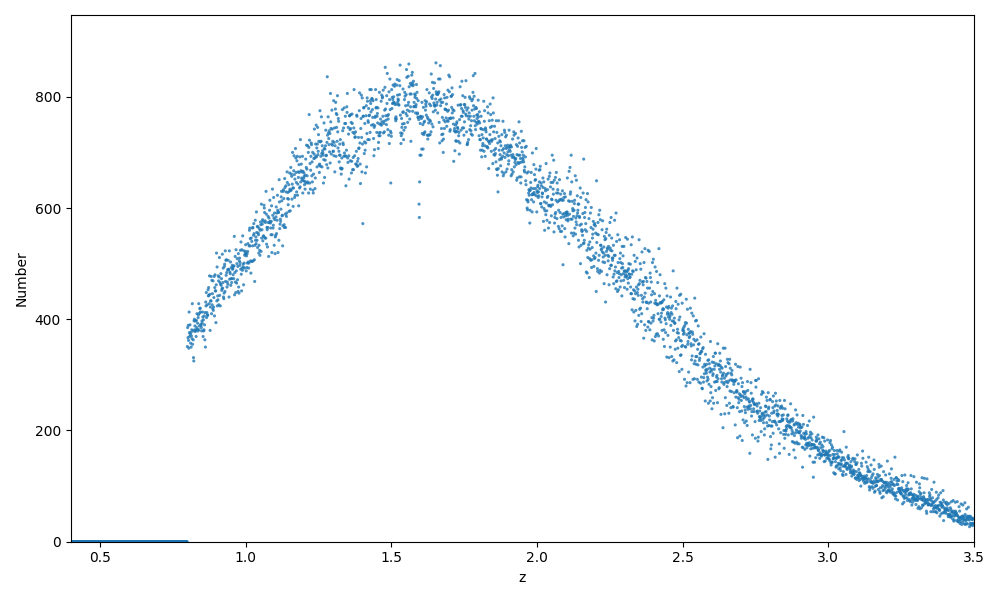}
    \caption{Redshift distributions of the QSO catalogues. Left to right: eBOSS DR14 (orange), eBOSS DR16 (green), and DESI DR1 (blue). The DR14 distribution is broad ($0.4 < z < 8$) but sparse at the extremes; we use the well-populated range $0.4 < z < 2.8$. The DESI DR1 distribution spans $0.8 < z < 3.5$; we use $0.8 < z < 3.0$ to avoid edge effects. All histograms use a bin width of $\Delta z = 0.001$. Note that the y-axis scales differ between panels. The primary overlap region for the eBOSS samples is $0.8 < z < 2.2$, while DESI significantly overlaps in the range $0.8 < z < 2.3$.}
    \label{fig:1}
\end{figure*}

\begin{figure*}[!htbp]
    \centering
    \includegraphics[width=6cm]{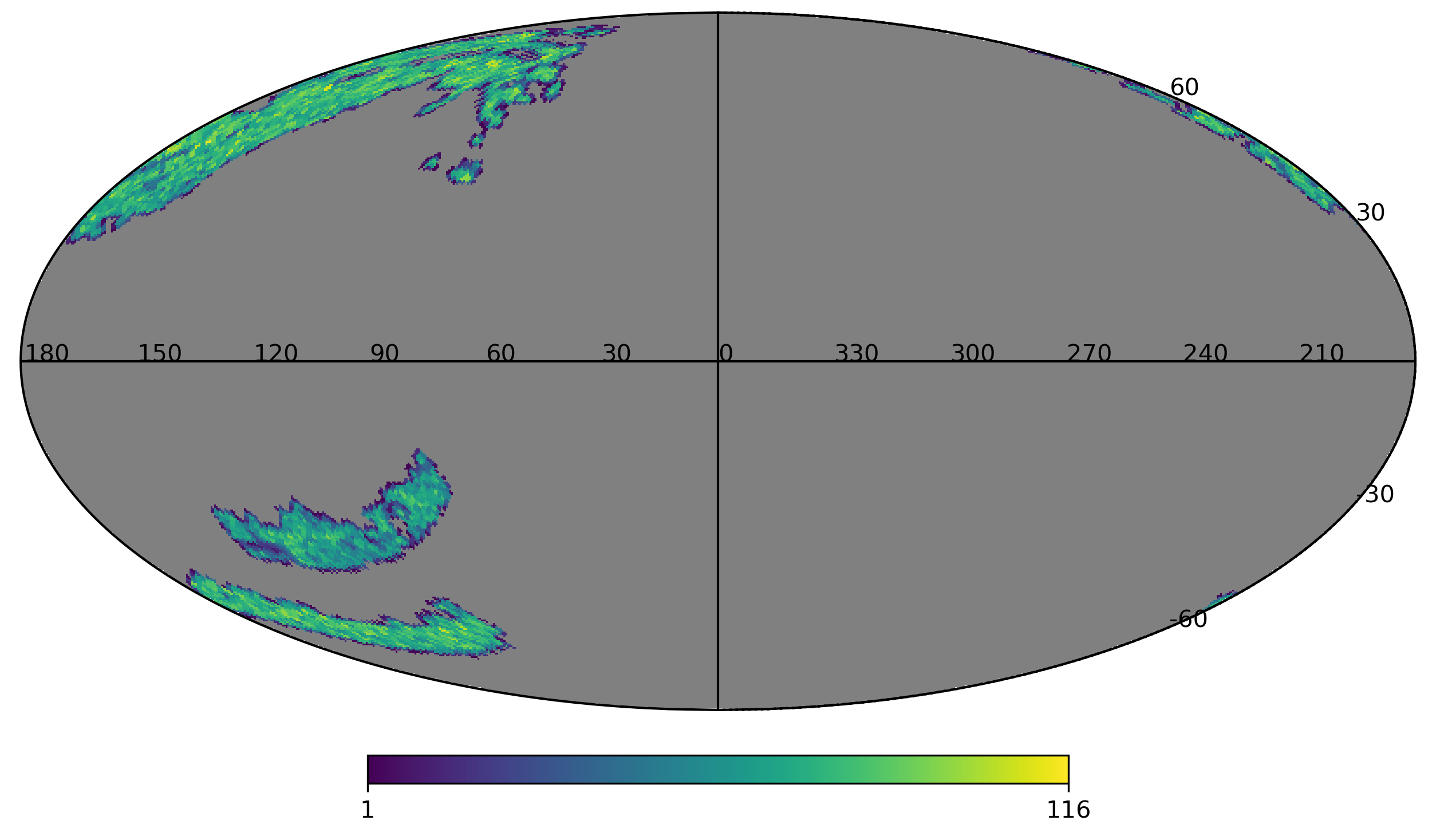}
    \includegraphics[width=6cm]{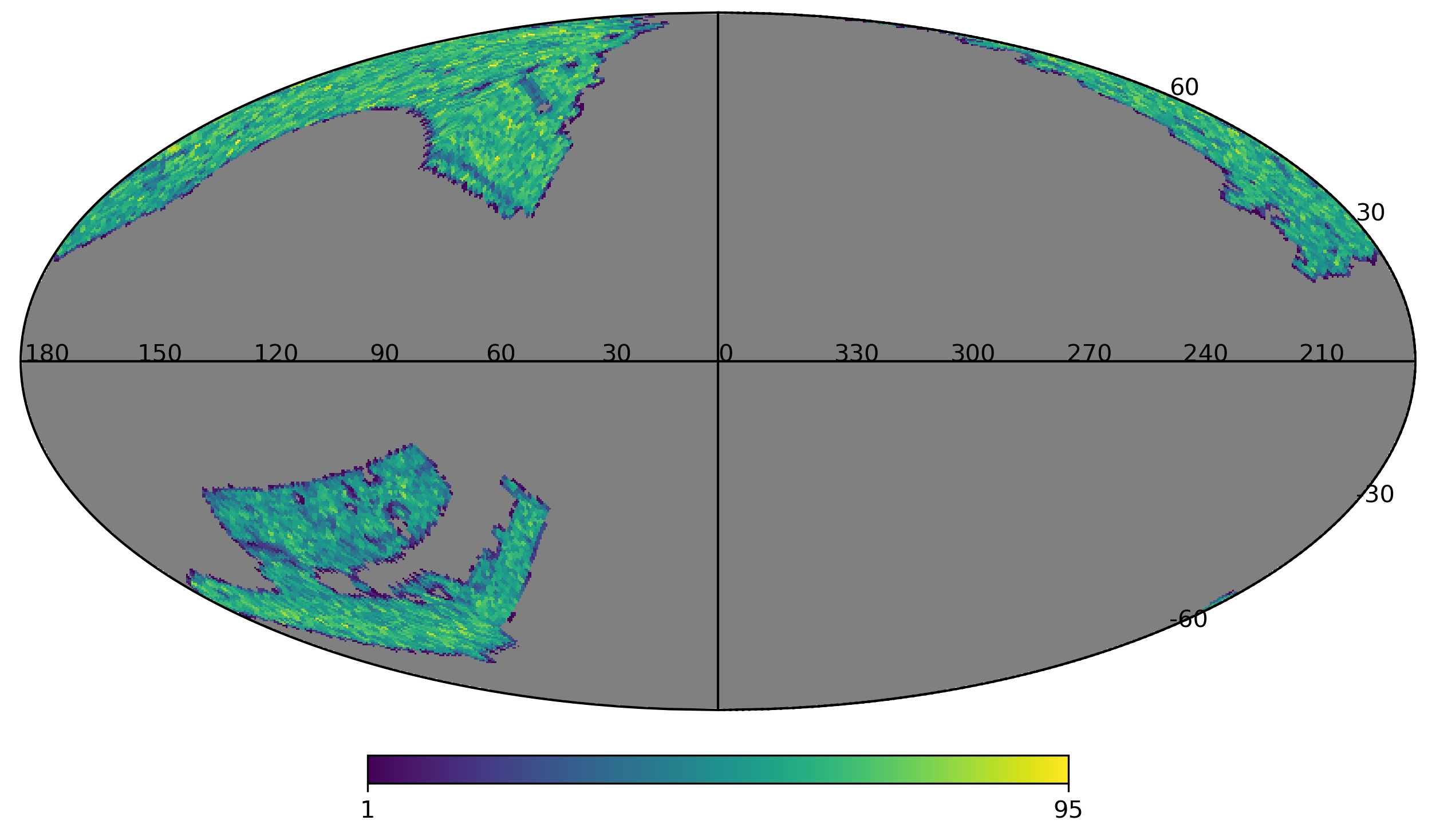}
    \includegraphics[width=6cm]{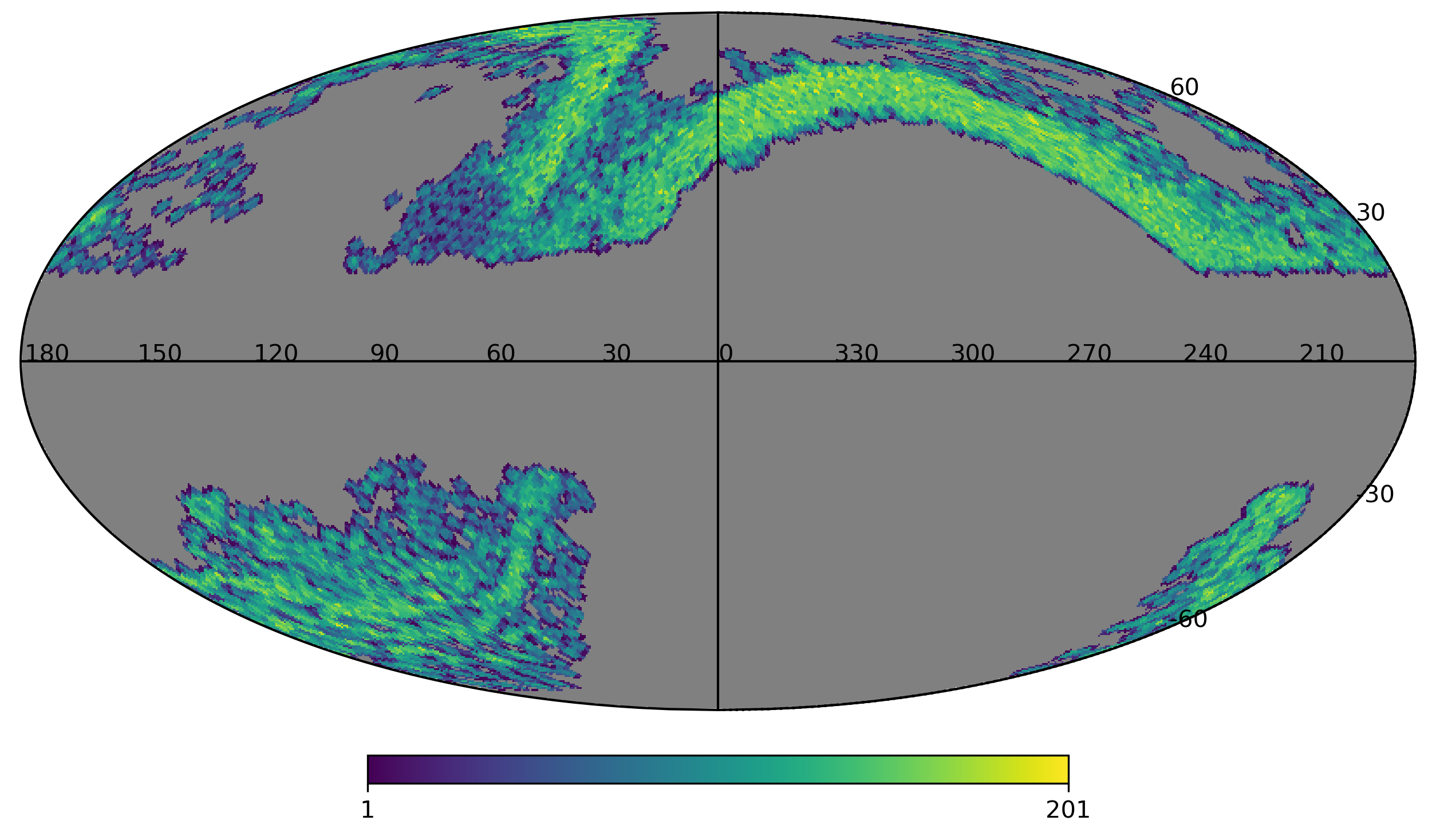}
    \caption{Sky footprints of the QSO samples in Mollweide projection (HEALPix format with $N_{\rm side}=64$). Left to right: eBOSS DR14, eBOSS DR16, and DESI DR1. The maps show the number count per pixel for each catalogue.}
    \label{fig:footprints}
\end{figure*}

The construction of this LSS catalogue, detailed in \citet{Adame_2025}, involves critical steps to correct for systematic variations in the observed number density and to ensure its suitability for cosmological clustering analyses, as follows.

Systematic weights: The catalogue incorporates weights to correct for two main sources of spurious density fluctuations: (1) imaging systematics (e.g. seeing, depth, and stellar density) affecting the input target selection, and (2) variations in the ability to confidently measure redshifts from DESI spectra, which depend on observing conditions and instrumental performance.

Reference randoms: A matched set of synthetic ‘randoms’ is provided to accurately model the survey selection function, geometry, and fibre assignment incompleteness.

Validation:The two-point clustering measurements from the final LSS catalogue have been compared to high-fidelity simulations of the DESI survey. They are found to be in general statistical agreement, typically within a 2$\%$ factor in the inferred galaxy bias, confirming that observational systematic uncertainties are sub-dominant to statistical uncertainties for large-scale analyses.

For our dipole analysis, we used the total systematic weight $w_i$ provided in the catalogue, which combines the corrections mentioned above. To avoid potential biases from sparse edges of the redshift distribution, we restricted our analysis to the well-characterized range $0.8 < z < 3.0$, yielding a final sample of 1,176,570 QSOs.

The redshift range and the total number of sources for each final sample are summarized in Table~\ref{tab:catalogs}. The redshift distributions are visualized in Figure~\ref{fig:1} as scatter plots with a bin width of $\Delta z = 0.001$

\begin{table}[!htbp]
\centering
\caption{QSO catalogues used in this analysis.}
\label{tab:catalogs}
\begin{tabular}{lccc}
\hline
Survey & Redshift Range & Objects  & Median $z$ \\
\hline
eBOSS DR14 QSO & $0.4 < z < 2.8$ & 187,954 & 1.59 \\
eBOSS DR16 QSO & $0.8 < z < 2.2$ & 343,708 & 1.51 \\
DESI DR1 QSO & $0.8 < z < 3.0$ & 1,176,570 & 1.75 \\
\hline
\end{tabular}
\tablefoot{The redshift ranges are selected to focus on the well-sampled core of each distribution while minimizing edge effects (see Figure~\ref{fig:1}).}
\end{table}

The sky footprints of the three catalogues are shown in Figure~\ref{fig:footprints}. To address potential hemispheric systematic biases from instrumentation or calibration, we performed a detailed comparison of the redshift distributions between the NGC and the SGC. As shown in Figure~\ref{fig:2}, we used bootstrap resampling to estimate the mean redshift in each cap. The NGC mean is taken as a reference (centred at zero). While the shapes of the NGC and SGC distributions are similar, they do not perfectly overlap. The systematic offset is approximately 0.008 for the eBOSS samples and 0.001 for the DESI samples, both of which are of the same order as the expected kinematic signal $\beta \equiv v/c \sim 0.001$. This significant offset could bias a combined NGC+SGC dipole estimate. Therefore, we implemented an unbiased estimation method, described in Section~\ref{'unbiased'}, which analyses the two hemispheres separately.

\section{Method}
\label{sec:method}

\subsection{Redshift dipole}
\label{sec:redshift dipole}

The kinematic redshift dipole arises from the Doppler shift due to our peculiar motion relative to the cosmic rest frame. In an isotropic universe, the observed redshift distribution of distant sources should exhibit a dipolar anisotropy modulated by this motion. The dominant contribution to the observed redshift dipole $\boldsymbol{\Delta}$ is our peculiar velocity $\boldsymbol{v}$, expressed as $\boldsymbol{\beta} = \boldsymbol{v}/c$, where $c$ is the speed of light. The theoretical relationship is given by \citet{Nadolny_2021}:
\begin{equation}
    \boldsymbol{\Delta} = -\boldsymbol{\beta} + \boldsymbol{\Delta}_{\rm cd} + \mathcal{O}(\beta^{2}),
\end{equation}
where $\boldsymbol{\Delta}_{\rm cd}$ is the clustering dipole from large-scale structure, and $\mathcal{O}(\beta^{2})$ represents higher-order terms

Due to our motion, the observed frequency ($\nu$) of light from a source in direction $\hat{\boldsymbol{n}}$ is related to its rest-frame frequency $\nu_{\rm rest}$ by
\begin{equation}
    \nu = \nu_{\rm rest} \, \delta(\hat{\boldsymbol{n}}, \boldsymbol{\beta}), \quad \text{where} \quad \delta(\hat{\boldsymbol{n}}, \boldsymbol{\beta}) = \frac{ 1 + \hat{\boldsymbol{n}} \cdot \boldsymbol{\beta} }{\sqrt{1 - \beta^2}}.
\end{equation}

Using $c = \lambda \nu$ and the definition of redshift $z = (\lambda_{\rm observe} - \lambda_{\rm emit}) / \lambda_{\rm emit}$, we derived the Lorentz boost relation between the observed redshift $z'$ (in the observer's frame) and the intrinsic redshift $z$ (in the CMB rest frame):
\begin{equation}
    1+z' = (1+z) \, \delta(\hat{\boldsymbol{n}}, \boldsymbol{\beta})^{-1}.
    \label{eq:boost}
\end{equation}

The term $z$ represents the expected redshift in an isotropic universe.

For high-redshift sources ($z \gtrsim 1$), such as the QSOs used here, the clustering dipole contribution $\boldsymbol{\Delta}_{\rm cd}$ is significantly suppressed compared to the kinematic term \citep{Tiwari_2016}. Therefore, to first approximation, we assumed $\boldsymbol{\Delta} \approx -\boldsymbol{\beta}$ when translating a measured dipole into a peculiar velocity.

\subsection{Estimator}
\label{sec:dipole estimator}

Following the tomographic methodology of \citet{da_Silveira_Ferreira_2024}, we estimated the kinematic dipole by analysing the redshift distribution in fine bins. This approach mitigates the impact of large-scale clustering noise and allows us to probe potential redshift evolution of the signal.

We divided the QSO sample into narrow redshift bins of width $\Delta z = 0.001$. For each bin, we modelled the observed redshift $z'_i$ as a Doppler-shifted version of an intrinsic, isotropic redshift field. The goal is to determine the dipole vector $\boldsymbol{\Delta} = (\Delta_x, \Delta_y, \Delta_z)$ defined in the Cartesian coordinate system transformed from Galactic coordinates ( $x$ towards the Galactic centre, $y$ in the direction of Galactic rotation, and $z$ towards the North Galactic Pole), and a monopole offset $m$ that best match this model. This is achieved by minimizing a per-bin $\chi^2$ statistic using a least-squares estimator. In practice, we employed the \texttt{fit\_dipole} function from the HEALPix package \citep{Gorski_2005}, which performs this fit on a spherical harmonic representation of the data within the bin.

The core of the estimator compares the observed redshifts with a simulated isotropic field. For a trial dipole $\boldsymbol{\Delta}$, we first computed the Lorentz boost factor $\delta(\hat{\boldsymbol{n}}_i, -\boldsymbol{\Delta})$ for each QSO $i$ in direction $\hat{\boldsymbol{n}}_i$. We then constructed a simulated intrinsic redshift for the bin, $z_{bin}^{\mathrm{simul}}$, by removing the estimated Doppler shift from all objects and computing the weighted mean:
\begin{equation}
    z_{bin}^{simul} = \frac{\sum_{i}^{N}w_{i}(1+z'_{i})\delta(\bm{\hat{n}_i},{-\bm{\Delta}})} {\sum_{i}^{N}w_{i}}-1.     
\end{equation}

This $z_{bin}^{\mathrm{simul}}$ represents our best estimate of what the redshift would be in the CMB rest frame for that bin, given the current $\boldsymbol{\Delta}$. The $N$ denotes the total number of objects. The total weights $w_i = w_{i,\mathrm{sys}} \cdot w_{i,\mathrm{fail}}$ are derived from the LSS catalogues, where $w_{i,\mathrm{sys}}$ represents the systematic weight and $w_{i,\mathrm{fail}}$ the redshift failure weight. We emphasize that the weight definition is consistent for both the eBOSS and DESI catalogues used in this analysis. Next, we applied the boost to this intrinsic value to predict what we should observe if $\boldsymbol{\Delta}$ were correct: ${z''_{i,bin} = (1+z_{bin}^{simul})\delta(\bm{\hat{n_i}},{-\bm\Delta})^{-1}-1.}$
The $\chi^2$ for the bin measures the discrepancy between the actual observations $z'_i$ and these predictions $z''_{i,bin}$:
\begin{equation}
    {\chi_{bin}^2(\bm\Delta) } =\frac{ {\textstyle \sum_{i}^{N}}w_{i}[1+z'_{i}-(1+z''_{i,bin})]^2}{\sum_{i}^{N}w_{i}}. 
\end{equation}

The estimator iteratively adjusts $\boldsymbol{\Delta}$ to minimize $\chi_{bin}^2$. The minimization process aims to fit a reference frame that is as stationary as possible relative to the CMB. This was achieved by removing our peculiar velocity. It should be noted that, since the peculiar velocity of the observed objects is unknown, we relied solely on the Earth's peculiar velocity and the statistical characteristics of the material reference frame for the fit. As mentioned in Section~\ref{sec:fisher}, the distant material reference frame should remain relatively stationary with respect to the CMB as a whole.

\subsection{Unbiased estimation}
\label{'unbiased'}

\begin{figure*}[!htbp]
    \centering
    \includegraphics[width=6cm]{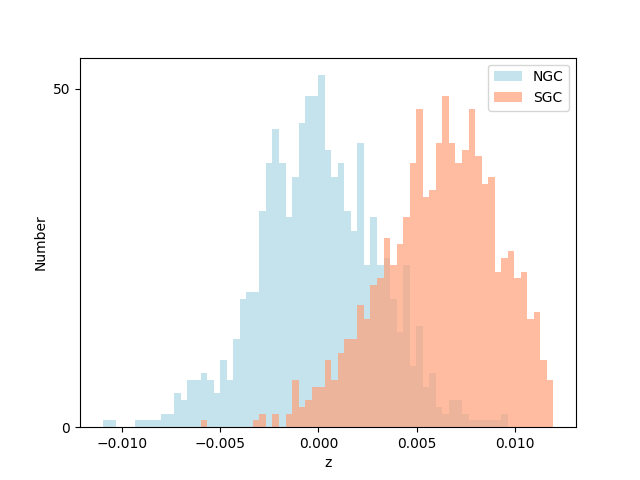}
    \includegraphics[width=6cm]{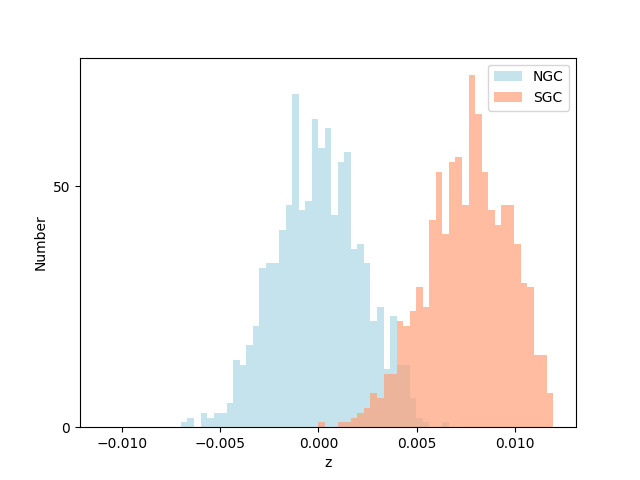}
    \includegraphics[width=6cm]{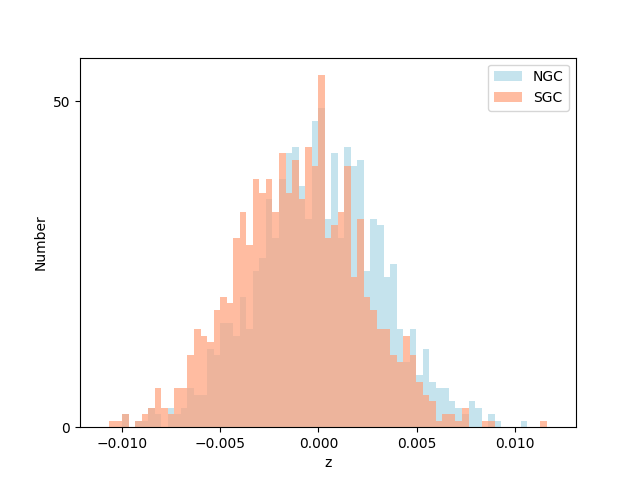}
    \caption{Assessment of NGC and SGC systematic differences via bootstrap resampling. For each QSO sample (left: eBOSS DR14; centre: eBOSS DR16; right: DESI DR1), we repeatedly (1000 times) drew bootstrap resamples of 28,000 data points from each hemisphere's dataset, computing the mean redshift for each resample. The distributions of these resampled means are plotted as histograms with a bin width of 0.000333. The y-axis represents the count of resamples whose mean redshift falls within the corresponding bin. All distributions are plotted relative to the average redshift of the total NGC dataset (vertical line at zero). The x-axis shows the redshift difference $\Delta z$ (dimensionless) relative to the total NGC mean. The offset between the NGC (centred) and SGC distributions indicates a hemisphere-level systematic discrepancy. The amplitude of this offset ($\sim 0.008$ for eBOSS, $\sim 0.001$ for DESI) is comparable to the expected kinematic dipole signal ($\beta = v/c \sim 10^{-3}$), necessitating separate hemispheric analysis (Section~\ref{'unbiased'}).}
    \label{fig:2}
\end{figure*}

As established in Section~\ref{sec:data} and Figure~\ref{fig:2}, significant systematic differences exist between the NGC and SGC. A joint fit to the combined data, assuming a single global monopole, would be biased because the mean redshift (the monopole) differs between hemispheres. To obtain an unbiased estimate of the dipole — which should be a global vector — we must account for these hemispheric offsets.

The key to our unbiased estimation is to perform the dipole fit separately on the NGC and SGC datasets while enforcing consistency in the resulting dipole vector. If the two Galactic caps are analysed together, the dipole calculation would be adjusted relative to the combined monopole of both hemispheres, introducing systematic differences of the same order—or larger—than the expected dipole signal (c$\Delta  \sim 300$ km/s). We therefore analysed the NGC and the SGC independently. 

In practice, within each redshift bin, we minimized $\chi^2_{\mathrm{bin,NGC}}(\boldsymbol{\Delta})$ and $\chi^2_{\mathrm{bin,SGC}}(\boldsymbol{\Delta})$, allowing the fit in each hemisphere to account for its own redshift weighting. Crucially, we then required that the dipole vector $\boldsymbol{\Delta}$ that minimizes the combined $\chi^2_{\mathrm{total}}$ in Equation~\eqref{eq:chi2_combined} be identical for both hemispheres. This approach effectively avoids the bias introduced by merging the data; it extracts a single, coherent kinematic dipole signal consistent across the full sky.

We implemented this by performing the minimization separately for the NGC and the SGC samples within each redshift bin. This yields two $\chi^2$ functions: $\chi^2_{\mathrm{bin,NGC}}(\boldsymbol{\Delta})$ and $\chi^2_{\mathrm{bin,SGC}}(\boldsymbol{\Delta})$, each already marginalized over their respective hemispheric monopoles. The total $\chi^2$ for the entire dataset is then the sum over all bins and both hemispheres:
\begin{equation}
\chi^2_{\mathrm{total}}(\boldsymbol{\Delta}) = \sum\left[ \chi^2_{\mathrm{bin,NGC}}(\boldsymbol{\Delta}) + \chi^2_{\mathrm{bin,SGC}}(\boldsymbol{\Delta}) \right].
\label{eq:chi2_combined}
\end{equation}
Minimizing $\chi^2_{\mathrm{total}}(\boldsymbol{\Delta})$ yields the dipole estimate ${\boldsymbol{\Delta}}$ that best fits the data while being insensitive to the NGC-SGC mean redshift offset.

\subsection{Fisher's statistical method}    
\label{sec:fisher}

To quantitatively assess the tension between our measured dipole and the prediction from the CMB kinematic dipole ($\boldsymbol{\Delta}_{\mathrm{CMB}}$), we needed a single statistical measure that combines the evidence from all three Cartesian components. Fisher's combined probability test \citep{FISHER1925} provides a rigorous framework for this.

For each Cartesian component $k \in {x, y, z}$, we defined a null hypothesis $H_0^{(k)}$: the measured dipole component $\Delta_k$ is consistent with the CMB-predicted value $\Delta_{\mathrm{CMB}, k}$. We computed the corresponding $p$-value, $P_k$, which represents the probability of obtaining a value as extreme as or more extreme than our measurement $\Delta_k$ if $H_0^{(k)}$ were true. Assuming Gaussian errors, $P_k$ is derived from the two-tailed test statistic $|\Delta_k - \Delta_{\mathrm{CMB}, k}| / \sigma$, where $\sigma = \sqrt{\sigma_{\Delta_k}^2 + \sigma_{\Delta_{\mathrm{CMB},k}}^2} \approx \sigma_{\Delta_k}$, since the CMB dipole is measured with high precision.

Following the CP and the work of \citet{10.1093/mnras/206.2.377}, the matter reference frame defined by distant objects is expected to be at rest relative to the CMB on large scales, meaning that an isotropic and homogeneous reference frame should not systematically bias any particular dipole component. Based on this, we assumed that the tests for the three orthogonal components are independent, Fisher's method combines the $p$-values into a single test statistic:
\begin{equation}
X^2 = -2 \sum_{k=1}^{3} \ln(P_k).
\label{eq:fisher_stat}
\end{equation}

If all individual null hypotheses are true and the test statistics $|\Delta_k - \Delta_{\mathrm{CMB}, k}|$ are independent, the test statistic $X^2$ follows a $\chi^2$ distribution with $2 \times 3 = 6$ degrees of freedom.

The final combined $p$-value, $P_{\mathrm{combined}}$, was calculated as the survival function of the $\chi^2_6$ distribution at the observed $X^2$. A very small $P_{\mathrm{combined}}$ indicates a low probability that all three CMB dipole component predictions are simultaneously consistent with our measurements. This combined $p$-value was then converted into a significance level in units of Gaussian standard deviations ($\sigma$). This metric provides a comprehensive and conservative measure of the overall discrepancy between the large-scale structure dipole and the CMB dipole.

\section{Discussion and conclusion}
\label{sec:concls}

\begin{figure*}[!htbp] 
    \centering
    \includegraphics[width=4.5cm]{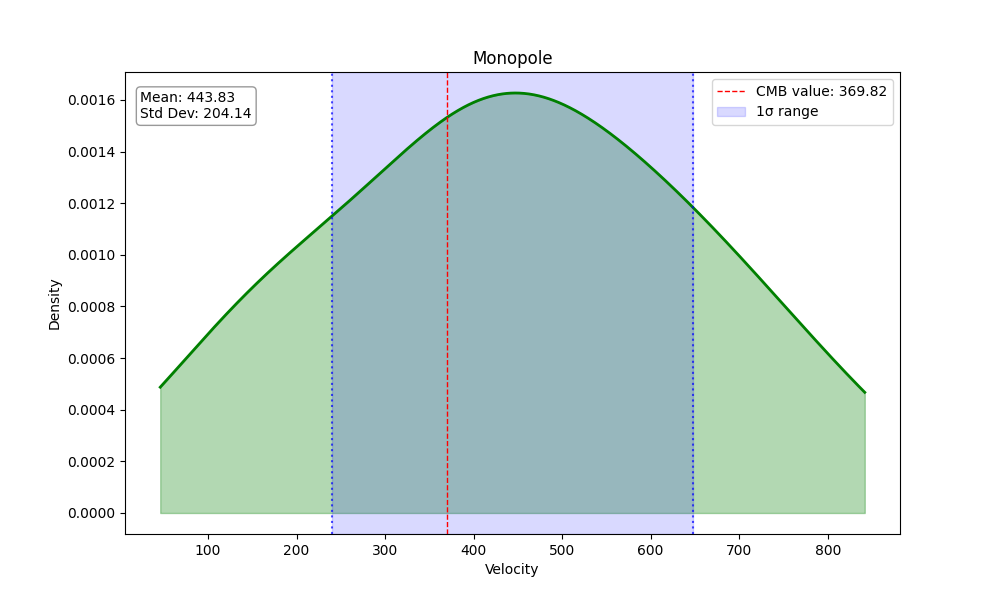}
    \includegraphics[width=4.5cm]{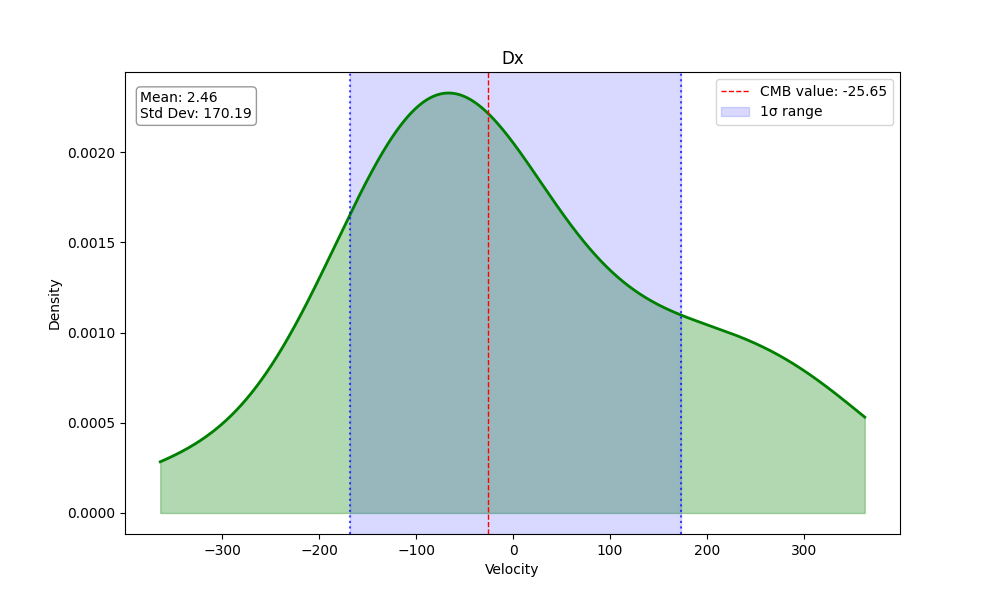}
    \includegraphics[width=4.5cm]{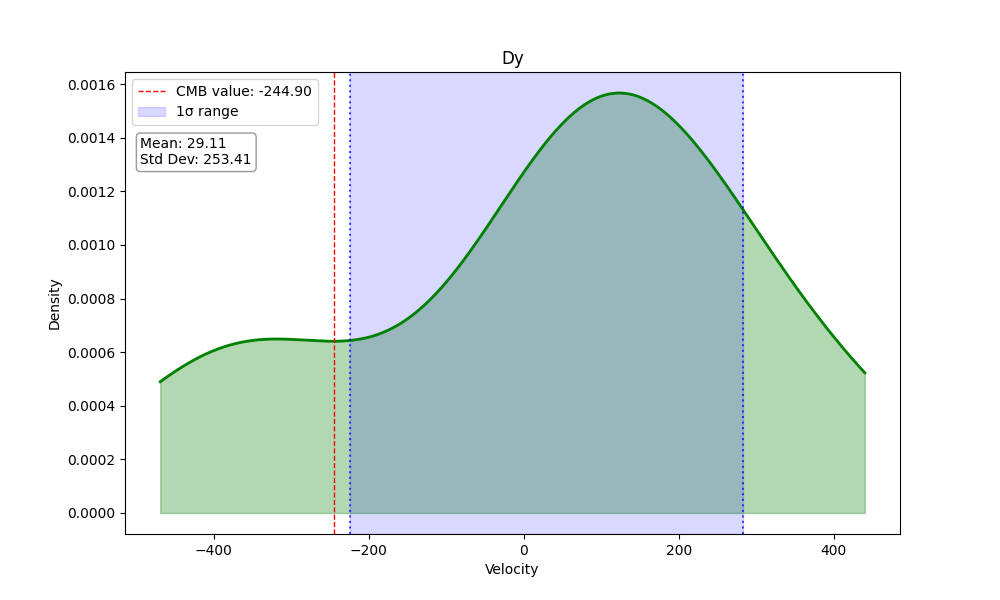}
    \includegraphics[width=4.5cm]{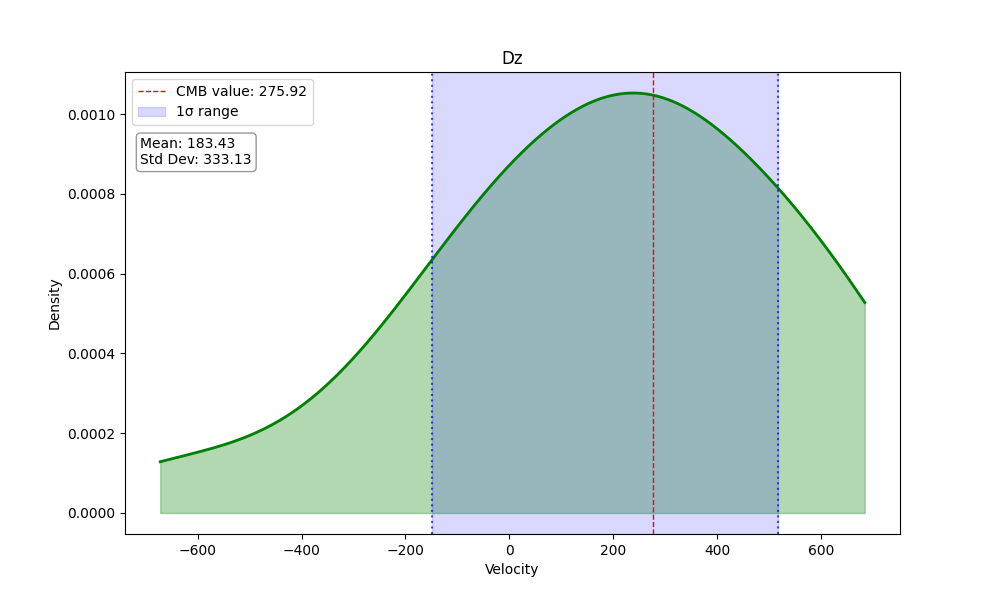}
    \caption{ Empirical distribution plots of the DESI DR1 redshift dipole in its Cartesian components (m, $c\Delta_x$, $c\Delta_y$, $c\Delta_z$). The red curve represents the dipole expected from the CMB dipole. The x-axis represents velocity (in kilometres per second), the y-axis represents distribution density, and the shaded regions denote the $1\sigma$ confidence intervals. Notably, the $c\Delta_y$ component shows small deviation from the CMB prediction, while the $1\sigma$ ranges of both $c\Delta_x$ and $c\Delta_z$ fully encompass the CMB dipole value.}
    \label{fig:3}
\end{figure*}

In this study, we presented a test of the CP by measuring the large-scale structure dipole anisotropy using QSO samples from eBOSS DR14 ($0.4 < z < 2.8$), DR16 ($0.6 < z < 2.2$), and DESI DR1 ($0.8 < z < 3.0$). Our analysis, employing the methodology of \citet{da_Silveira_Ferreira_2024}, allows for a direct comparison of the kinematic dipole inferred from the distribution of QSOs with the canonical CMB dipole. The key results, summarized in Table~\ref{tab:2} and Fig.~\ref{fig:3}, reveal a complex picture that warrants careful scientific discussion.

From the DESI DR1 data, we infer a bulk flow velocity of $v = 443.8 \pm 204.1$ km/s towards $(l, b) = (107.4^\circ \pm 86.8^\circ, 28.4^\circ \pm 45.2^\circ)$. While the velocity magnitude is consistent with the CMB kinematic value of 370 km/s within $1.56\sigma$, the best-fit direction shows a substantial offset from the CMB dipole $(l = 264.02^\circ, b = 48.25^\circ)$. However, a component-by-component statistical assessment using Fisher's method yields an overall tension of $3.45\sigma$ for DESI, primarily driven by a deviation in the $\Delta_y$ component. To interpret these seemingly contradictory signals - a consistent velocity magnitude but offset $\Delta_{y}$ direction and component level tension - we considered the evidence in a broader context.

Internal consistency and data evolution: The progression from the highly anomalous eBOSS DR14 result to the CMB-consistent eBOSS DR16 result underscores the critical impact of data quality. The DR14 discrepancy is likely attributable to its smaller sample size and higher redshift dispersion, limitations mitigated in DR16. This evolution cautions against over-interpreting significant tensions from any single data release, including the $3.45\sigma$ value from DESI DR1

Methodological validation and external comparison: Our successful replication of the \citet{da_Silveira_Ferreira_2024} analysis on the eBOSS DR16 QSO sample yields a consistent result. We derive a bulk flow of $v = 477.6 \pm 224.4$ km/s towards $(l, b) = (261.4^\circ \pm 117.3^\circ, -0.4^\circ \pm 34.6^\circ)$, in agreement with the CMB kinematic dipole at a $2.1\sigma$ level (see Fig.~\ref{appfig1}). This is in close agreement with their original finding of $v = 196_{-79}^{+92}$ km/s towards $(l, b) = (298_{~~-52}^{\circ+29}, 50_{~~-62}^{\circ+8})$, which they reported as consistent with the CMB kinematic expectation at a $2\sigma$ level. This independent verification validates our analytical framework and, crucially, confirms that the methodology itself does not inherently produce high-tension results. The discrepancy, therefore, arises specifically with the DESI DR1 sample, suggesting sample-specific factors may be at play.

The nature of the $3.45\sigma$ tension in DESI: The significant Fisher combined tension for DESI warrants careful dissection. As Fig.~\ref{fig:3} shows, it stems largely from the $\Delta_y$ component, whose associated unipolar amplitude differs from the expected value by nearly 300 km/s. This deviation may originate from the data itself, potentially challenging the CP. This pattern—a component-isolated deviation—is more indicative of residual systematic effects or unusual sample variance within DESI's specific scan pattern and calibration, rather than a coherent, all-sky anisotropy expected from a genuine breakdown of isotropy.

The central question is whether the measured dipole reflects our motion relative to the Hubble flow (kinematic origin, supporting CP) or an intrinsic anisotropy (challenging CP). The amplitude of the DESI-inferred bulk flow is consistent with the kinematic expectation, which is a necessary condition for the kinematic interpretation. However, the offset in the best-fit direction, though not statistically significant given the large uncertainties, introduces ambiguity. Claims of a CP violation require not only high statistical significance but also robustness across multiple independent probes, elimination of plausible systematics, and a consistent directional signal across different redshift slices and tracers. Our DESI result, characterized by a component-level tension and a direction that is not robustly constrained, currently falls short of these stringent criteria.

In conclusion, our analysis employed the inferred bulk flow velocity—derived from the dipole—as the critical test. The amplitude of this velocity, at $|v| = 443.8 \pm 204.1$ km/s, is consistent with the kinematic interpretation of the CMB dipole within $1.56\sigma$. However, its direction $(l, b) = (107.4^\circ \pm 86.8^\circ, 28.4^\circ \pm 45.2^\circ)$ shows a notable offset from the CMB dipole, though with large uncertainties. The significant statistical tension ($3.45\sigma$) is primarily localized to the $\Delta_y$ component, which exhibits a large deviation. Crucially, the $\Delta_x$ and $\Delta_z$ components are in good agreement with expectations (see Fig.~\ref{fig:3}). This pattern—a coherent amplitude, a directionally ambiguous vector with two well-behaved components—suggests that the tension is more likely driven by residual systematics or sample variance in DESI DR1, rather than by cosmological anisotropy. Therefore, the current measurements do not provide compelling evidence against the CP. Future, larger datasets will be essential to reduce uncertainties and clarify the origin of the $\Delta_y$ anomaly.

\begin{table}[!htbp]
	\centering
        \small 
        \setlength{\tabcolsep}{3pt} 
	\caption{Three QSO catalogues dipole result, with 1$\sigma$ range.}
	\begin{tabular}{c | c c c c}
		\hline
		\hline
		Case & $z$ & $\left |v  \right |$ (km/s) & $l$(deg) & $b$(deg) \\
		\hline
        CMB & 1090 & $369.8\pm0.01$ & $264.02\pm0.01$ & $48.253\pm0.01$\\
		eBOSS DR14 & 1.59 & $820.9\pm209.5$ & $104.2\pm73.3$ & $-17.2\pm37.8$ \\
		eBOSS DR16 & 1.51 & $477.6\pm224.4$ & $261.4\pm117.3$ & $-0.4\pm34.6$\\
        DESI DR1 & 1.75 & $443.8\pm204.1$ & $107.4\pm86.8$ & $28.4\pm45.2$ \\
		\hline
	\end{tabular} \label{tab:2}
    \tablefoot{We can obvserve the Dr14 dipole has a significant deviation from the CMB dipole and the Dr16 and DESI dipole is close to the CMB dipole.}
\end{table}

\begin{acknowledgements}
This work was supported by the National SKA Program of China (Grants Nos. 2022SKA0110200 and 2022SKA0110203). This work was also supported by Xiaofeng Yang's ZHISHAN Distinguished Professor startup funding of Henan University.
\end{acknowledgements}

\section*{Data availability}

The eBOSS data are available at~\url{https://data.sdss.org/sas/dr16/eboss}.
The DESI data are available at~\url{https://data.desi.lbl.gov/public/dr1}.
    
\bibliographystyle{aa.bst}
\bibliography{reference.bib}

\begin{thebibliography}{41}
\expandafter\ifx\csname natexlab\endcsname\relax\def\natexlab#1{#1}\fi

\bibitem[{Abolfathi {et~al.}(2018)Abolfathi, Aguado, Aguilar, Prieto, Almeida, Ananna, Anders, Anderson, Andrews, Anguiano, Aragón-Salamanca, Argudo-Fernández, Armengaud, Ata, Aubourg, Avila-Reese, Badenes, Bailey, Balland, Barger, Barrera-Ballesteros, Bartosz, Bastien, Bates, Baumgarten, Bautista, Beaton, Beers, Belfiore, Bender, Bernardi, Bershady, Beutler, Bird, Bizyaev, Blanc, Blanton, Blomqvist, Bolton, Boquien, Borissova, Bovy, Bradna~Diaz, Nielsen~Brandt, Brinkmann, Brownstein, Bundy, Burgasser, Burtin, Busca, Cañas, Cano-Díaz, Cappellari, Carrera, Casey, Sodi, Chen, Cherinka, Chiappini, Choi, Chojnowski, Chuang, Chung, Clerc, Cohen, Comerford, Comparat, do~Nascimento, da~Costa, Cousinou, Covey, Crane, Cruz-Gonzalez, Cunha, Ilha, Damke, Darling, Davidson, Dawson, de~Icaza~Lizaola, Macorra, de~la Torre, De~Lee, Sainte~Agathe, Deconto~Machado, Dell’Agli, Delubac, Diamond-Stanic, Donor, Downes, Drory, Mas~des Bourboux, Duckworth, Dwelly, Dyer, Ebelke, Eigenbrot, Eisenstein, Elsworth, Emsellem,
  Eracleous, Erfanianfar, Escoffier, Fan, Alvar, Fernandez-Trincado, Cirolini, Feuillet, Finoguenov, Fleming, Font-Ribera, Freischlad, Frinchaboy, Fu, Chew, Galbany, García~Pérez, Garcia-Dias, García-Hernández, Garma~Oehmichen, Gaulme, Gelfand, Gil-Marín, Gillespie, Goddard, González~Hernández, Gonzalez-Perez, Grabowski, Green, Grier, Gueguen, Guo, Guy, Hagen, Hall, Harding, Hasselquist, Hawley, Hayes, Hearty, Hekker, Hernandez, Hernandez~Toledo, Hogg, Holley-Bockelmann, Holtzman, Hou, Hsieh, Hunt, Hutchinson, Hwang, Jimenez~Angel, Johnson, Jones, Jönsson, Jullo, Sakil~Khan, Kinemuchi, Kirkby, Kirkpatrick~IV, Kitaura, Knapp, Kneib, Kollmeier, Lacerna, Lane, Lang, Law, Le~Goff, Lee, Li, Li, Lian, Liang, Lima, Lin~林俐, Long, Lucatello, Lundgren, Mackereth, MacLeod, Mahadevan, Geimba~Maia, Majewski, Manchado, Maraston, Mariappan, Marques-Chaves, Masseron, Masters~何凱, McDermid, McGreer, Melendez, Meneses-Goytia, Merloni, Merrifield, Meszaros, Meza, Minchev, Minniti, Mueller, Muller-Sanchez, Muna,
  Muñoz, Myers, Nair, Nandra, Ness, Newman, Nichol, Nidever, Nitschelm, Noterdaeme, O’Connell, Oelkers, Oravetz, Oravetz, Ortíz, Osorio, Pace, Padilla, Palanque-Delabrouille, Palicio, Pan, Pan, Parikh, Pâris, Park, Peirani, Pellejero-Ibanez, Penny, Percival, Perez-Fournon, Petitjean, Pieri, Pinsonneault, Pisani, Prada, Prakash, de~Andrade~Queiroz, Raddick, Raichoor, Rembold, Richstein, Riffel, Riffel, Rix, Robin, Torres, Román-Zúñiga, Ross, Rossi, Ruan, Ruggeri, Ruiz, Salvato, Sánchez, Sánchez, Almeida, Sánchez-Gallego, Rojas, Santiago, Schiavon, Schimoia, Schlafly, Schlegel, Schneider, Schuster, Schwope, Seo, Serenelli, Shen, Shen, Shetrone, Shull, Aguirre, Simon, Skrutskie, Slosar, Smethurst, Smith, Sobeck, Somers, Souter, Souto, Spindler, Stark, Stassun, Steinmetz, Stello, Storchi-Bergmann, Streblyanska, Stringfellow, Suárez, Sun, Szigeti, Taghizadeh-Popp, Talbot, Tang, Tao, Tayar, Tembe, Teske, Thakar, Thomas, Tissera, Tojeiro, Tremonti, Troup, Urry, Valenzuela, Bosch, Vargas-González,
  Vargas-Magaña, Vazquez, Villanova, Vogt, Wake, Wang, Weaver, Weijmans, Weinberg, Westfall, Whelan, Wilcots, Wild, Williams, Wilson, Wood-Vasey, Wylezalek, Xiao~肖, Yan, Yang, Ybarra, Yèche, Zakamska, Zamora, Zarrouk, Zasowski, Zhang, Zhao, Zhao, Zheng, Zheng, Zhou, Zhu, Zinn, \& Zou}]{Abolfathi_2018}
Abolfathi, B., Aguado, D.~S., Aguilar, G., {et~al.} 2018, \apjs, 235, 42

\bibitem[{Adame {et~al.}(2025)Adame, Aguilar, Ahlen, Alam, Alexander, Alvarez, Alves, Anand, Andrade, Armengaud, Avila, Aviles, Awan, Bailey, Baltay, Bault, Behera, BenZvi, Beutler, Bianchi, Blake, Blum, Brieden, Brodzeller, Brooks, Brown, Buckley-Geer, Burtin, Calderon, Canning, Carnero~Rosell, Cereskaite, Cervantes-Cota, Chabanier, Chaussidon, Chaves-Montero, Chen, Chen, Claybaugh, Cole, Cuceu, Davis, Dawson, de~la Macorra, de~Mattia, Deiosso, Demina, Dey, Dey, Ding, Doel, Edelstein, Eftekharzadeh, Eisenstein, Elliott, Fagrelius, Fanning, Ferraro, Ereza, Findlay, Flaugher, Font-Ribera, Forero-Sánchez, Forero-Romero, Frenk, Garcia-Quintero, Gaztañaga, Gil-Marín, Gontcho, Gonzalez-Morales, Gonzalez-Perez, Gordon, Green, Gruen, Gsponer, Gutierrez, Guy, Hadzhiyska, Hahn, Hanif, Herrera-Alcantar, Honscheid, Hou, Howlett, Huterer, Iršič, Ishak, Juneau, Karaçaylı, Kehoe, Kent, Kirkby, Kitaura, Kong, Kremin, Krolewski, Lai, Lan, Landriau, Lang, Lasker, Le~Goff, Le~Guillou, Leauthaud, Levi, Li, Lodha,
  Magneville, Manera, Margala, Martini, Maus, McDonald, Medina-Varela, Meisner, Mena-Fernández, Miquel, Moon, Moore, Moustakas, Mudur, Mueller, Muñoz-Gutiérrez, Myers, Nadathur, Napolitano, Neveux, Newman, Nguyen, Nie, Niz, Noriega, Padmanabhan, Paillas, Palanque-Delabrouille, Pan, Penmetsa, Percival, Pieri, Pinon, Poppett, Porredon, Prada, Pérez-Fernández, Pérez-Ràfols, Rabinowitz, Raichoor, Ramírez-Pérez, Ramirez-Solano, Rashkovetskyi, Ravoux, Rezaie, Rich, Rocher, Rockosi, Roe, Rosado-Marin, Ross, Rossi, Ruggeri, Ruhlmann-Kleider, Samushia, Sanchez, Saulder, Schlafly, Schlegel, Scholte, Schubnell, Seo, Sharples, Silber, Slosar, Smith, Sprayberry, Tan, Tarlé, Trusov, Vaisakh, Valcin, Valdes, Vargas-Magaña, Verde, Walther, Wang, Wang, Weaver, Weaverdyck, Wechsler, Weinberg, White, Wilson, Yu, Yu, Yuan, Yèche, Zaborowski, Zarrouk, Zhang, Zhao, Zhao, Zhou, \& Zou}]{Adame_2025}
Adame, A., Aguilar, J., Ahlen, S., {et~al.} 2025, \jcap, 2025, 017

\bibitem[{Aghanim {et~al.}(2020)Aghanim, Akrami, Arroja, Ashdown, Aumont, Baccigalupi, Ballardini, Banday, Barreiro, Bartolo, Basak, Battye, Benabed, Bernard, Bersanelli, Bielewicz, Bock, Bond, Borrill, Bouchet, Boulanger, Bucher, Burigana, Butler, Calabrese, Cardoso, Carron, Casaponsa, Challinor, Chiang, Colombo, Combet, Contreras, Crill, Cuttaia, de~Bernardis, de~Zotti, Delabrouille, Delouis, Désert, Di~Valentino, Dickinson, Diego, Donzelli, Doré, Douspis, Ducout, Dupac, Efstathiou, Elsner, Enßlin, Eriksen, Falgarone, Fantaye, Fergusson, Fernandez-Cobos, Finelli, Forastieri, Frailis, Franceschi, Frolov, Galeotta, Galli, Ganga, Génova-Santos, Gerbino, Ghosh, González-Nuevo, Górski, Gratton, Gruppuso, Gudmundsson, Hamann, Handley, Hansen, Helou, Herranz, Hildebrandt, Hivon, Huang, Jaffe, Jones, Karakci, Keihänen, Keskitalo, Kiiveri, Kim, Kisner, Knox, Krachmalnicoff, Kunz, Kurki-Suonio, Lagache, Lamarre, Langer, Lasenby, Lattanzi, Lawrence, Le~Jeune, Leahy, Lesgourgues, Levrier, Lewis, Liguori, Lilje,
  Lilley, Lindholm, López-Caniego, Lubin, Ma, Macías-Pérez, Maggio, Maino, Mandolesi, Mangilli, Marcos-Caballero, Maris, Martin, Martinelli, Martínez-González, Matarrese, Mauri, McEwen, Meerburg, Meinhold, Melchiorri, Mennella, Migliaccio, Millea, Mitra, Miville-Deschênes, Molinari, Moneti, Montier, Morgante, Moss, Mottet, Münchmeyer, Natoli, Nørgaard-Nielsen, Oxborrow, Pagano, Paoletti, Partridge, Patanchon, Pearson, Peel, Peiris, Perrotta, Pettorino, Piacentini, Polastri, Polenta, Puget, Rachen, Reinecke, Remazeilles, Renault, Renzi, Rocha, Rosset, Roudier, Rubiño-Martín, Ruiz-Granados, Salvati, Sandri, Savelainen, Scott, Shellard, Shiraishi, Sirignano, Sirri, Spencer, Sunyaev, Suur-Uski, Tauber, Tavagnacco, Tenti, Terenzi, Toffolatti, Tomasi, Trombetti, Valiviita, Van~Tent, Vibert, Vielva, Villa, Vittorio, Wandelt, Wehus, White, White, Zacchei, \& Zonca}]{2020}
Aghanim, N., Akrami, Y., Arroja, F., {et~al.} 2020, \aap, 641, A1

\bibitem[{Bengaly {et~al.}(2016)Bengaly, Bernui, Ferreira, \& Alcaniz}]{Bengaly_2016}
Bengaly, C. A.~P., Bernui, A., Ferreira, I.~S., \& Alcaniz, J.~S. 2016, \mnras, 466, 2799–2804

\bibitem[{{Blake} \& {Wall}(2002)}]{2002Natur.416..150B}
{Blake}, C. \& {Wall}, J. 2002, Nature, 416, 150

\bibitem[{Böhme {et~al.}(2025)Böhme, Schwarz, Tiwari, Pashapour-Ahmadabadi, Bahr-Kalus, Bilicki, Hale, Heneka, \& Siewert}]{B_hme_2025}
Böhme, L., Schwarz, D.~J., Tiwari, P., {et~al.} 2025, \prl, 135

\bibitem[{Clowes {et~al.}(2013)Clowes, Harris, Raghunathan, Campusano, Söchting, \& Graham}]{10.1093/mnras/sts497}
Clowes, R.~G., Harris, K.~A., Raghunathan, S., {et~al.} 2013, \mnras, 429, 2910

\bibitem[{Collaboration {et~al.}(2025)Collaboration, Abdul-Karim, Adame, Aguado, Aguilar, Ahlen, Alam, Aldering, Alexander, Alfarsy, Allen, Prieto, Alves, Anand, Andrade, Armengaud, Avila, Aviles, Awan, Bailey, Lizancos, Ballester, Bault, Bautista, BenZvi, e~Silva, Bermejo-Climent, Beutler, Bianchi, Blake, Blum, Bolton, Bonici, Brieden, Brodzeller, Brooks, Buckley-Geer, Burtin, Canning, Rosell, Carr, Carrilho, Casas, Castander, Cereskaite, Cervantes-Cota, Chaussidon, Chaves-Montero, Chen, Chen, Claybaugh, Cole, Cooper, Cousinou, Cuceu, Davis, Dawson, de~Belsunce, de~la Cruz, de~la Macorra, de~Mattia, Deiosso, Costa, Demina, Demirbozan, DeRose, Dey, Dey, Ding, Ding, Doel, Douglass, Dowicz, Ebina, Edelstein, Eisenstein, Elbers, Emas, Escoffier, Fagrelius, Fan, Fanning, Fawcett, Fernández-García, Ferraro, Findlay, Font-Ribera, Forero-Romero, Forero-Sánchez, Frenk, Gänsicke, Galbany, García-Bellido, Garcia-Quintero, Garrison, Gaztañaga, Gil-Marín, Gnedin, Gontcho, Gonzalez-Morales, Gonzalez-Perez, Gordon,
  Graur, Green, Gruen, Gsponer, Guandalin, Gutierrez, Guy, Hahn, Han, Han, He, Herrera-Alcantar, Honscheid, Hou, Howlett, Huterer, Iršič, Ishak, Jacques, Jimenez, Jing, Joachimi, Joudaki, Joyce, Jullo, Juneau, Karaçaylı, Karim, Kehoe, Kent, Khederlarian, Kirkby, Kisner, Kitaura, Kizhuprakkat, Kong, Koposov, Kremin, Krolewski, Lahav, Lai, Lamman, Lan, Landriau, Lang, Lange, Lasker, Goff, Guillou, Leauthaud, Levi, Li, Li, Lodha, Lokken, Luo, Magneville, Manera, Manser, Margala, Martini, Maus, McCullough, McDonald, Medina, Medina-Varela, Meisner, Mena-Fernández, Menegas, Mezcua, Miquel, Montero-Camacho, Moon, Moustakas, Muñoz-Gutiérrez, Muñoz-Santos, Myers, Myles, Nadathur, Najita, Napolitano, Newman, Nikakhtar, Nikutta, Niz, Noriega, Padmanabhan, Paillas, Palanque-Delabrouille, Palmese, Pan, Pan, Parkinson, Peacock, Percival, Pérez-Fernández, Pérez-Ràfols, Peterson, Piat, Pieri, Pinon, Poppett, Porredon, Prada, Pucha, Qin, Rabinowitz, Raichoor, Ramírez-Pérez, Ramirez-Solano, Rashkovetskyi, Ravoux,
  Riley, Rocher, Rockosi, Rohlf, Ross, Rossi, Ruggeri, Ruhlmann-Kleider, Sabiu, Said, Saintonge, Samushia, Sanchez, Sanders, Saulder, Schlafly, Schlegel, Scholte, Schubnell, Seo, Shafieloo, Sharples, Silber, Siudek, Smith, Sprayberry, Suárez-Pérez, Swanson, Tan, Tarlé, Taylor, Thomas, Tojeiro, Turner, Turner, Ureña-López, Vaisakh, Valluri, Vargas-Magaña, Verde, Walther, Wang, Wang, Wang, Weaver, Weaverdyck, Wechsler, White, Wolfson, Yang, Yèche, Youles, Yu, Yuan, Zaborowski, Zarrouk, Zhang, Zhao, Zhao, Zheng, Zhou, Zou, Zou, \& Zu}]{desicollaboration2025datarelease1dark}
Collaboration, D., Abdul-Karim, M., Adame, A.~G., {et~al.} 2025, Data Release 1 of the Dark Energy Spectroscopic Instrument

\bibitem[{da~Silveira~Ferreira \& Marra(2024)}]{da_Silveira_Ferreira_2024}
da~Silveira~Ferreira, P. \& Marra, V. 2024, \jcap, 2024, 077

\bibitem[{Dam {et~al.}(2023)Dam, Lewis, \& Brewer}]{Dam_2023}
Dam, L., Lewis, G.~F., \& Brewer, B.~J. 2023, \mnras, 525, 231–245

\bibitem[{Ellis \& Baldwin(1984)}]{10.1093/mnras/206.2.377}
Ellis, G. F.~R. \& Baldwin, J.~E. 1984, \mnras, 206, 377

\bibitem[{Eriksen {et~al.}(2004)Eriksen, Hansen, Banday, Gorski, \& Lilje}]{Eriksen_2004}
Eriksen, H.~K., Hansen, F.~K., Banday, A.~J., Gorski, K.~M., \& Lilje, P.~B. 2004, \apj, 605, 14–20

\bibitem[{Fisher(1925)}]{FISHER1925}
Fisher, R, A. 1925, Oliver and Boyd

\bibitem[{Franco {et~al.}(2023)Franco, Avila, \& Bernui}]{Franco_2023}
Franco, C., Avila, F., \& Bernui, A. 2023, \mnras, 527, 7400–7413

\bibitem[{Friday {et~al.}(2022)Friday, Clowes, \& Williger}]{10.1093/mnras/stac269}
Friday, T., Clowes, R.~G., \& Williger, G.~M. 2022, \mnras, 511, 4159

\bibitem[{Gibelyou \& Huterer(2012)}]{Gibelyou_2012}
Gibelyou, C. \& Huterer, D. 2012, \mnras, 427, 1994–2021

\bibitem[{Gorski {et~al.}(2005)Gorski, Hivon, Banday, Wandelt, Hansen, Reinecke, \& Bartelmann}]{Gorski_2005}
Gorski, K.~M., Hivon, E., Banday, A.~J., {et~al.} 2005, \apj, 622, 759–771

\bibitem[{Gupta \& Saini(2010)}]{Gupta_2010}
Gupta, S. \& Saini, T.~D. 2010, \mnras, 407, 651–657

\bibitem[{Haslbauer {et~al.}(2020)Haslbauer, Banik, \& Kroupa}]{Haslbauer_2020}
Haslbauer, M., Banik, I., \& Kroupa, P. 2020, \mnras, 499, 2845–2883

\bibitem[{Hinshaw {et~al.}(2009)Hinshaw, Weiland, Hill, Odegard, Larson, Bennett, Dunkley, Gold, Greason, Jarosik, Komatsu, Nolta, Page, Spergel, Wollack, Halpern, Kogut, Limon, Meyer, Tucker, \& Wright}]{Hinshaw_2009}
Hinshaw, G., Weiland, J.~L., Hill, R.~S., {et~al.} 2009, \apjs, 180, 225–245

\bibitem[{Keenan {et~al.}(2013)Keenan, Barger, \& Cowie}]{Keenan_2013}
Keenan, R.~C., Barger, A.~J., \& Cowie, L.~L. 2013, \apj, 775, 62

\bibitem[{{Lyke} {et~al.}(2020){Lyke}, {Higley}, {McLane}, {Schurhammer}, {Myers}, {Ross}, {Dawson}, {Chabanier}, {Martini}, {Busca}, {Mas des Bourboux}, {Salvato}, {Streblyanska}, {Zarrouk}, {Burtin}, {Anderson}, {Bautista}, {Bizyaev}, {Brandt}, {Brinkmann}, {Brownstein}, {Comparat}, {Green}, {de la Macorra}, {Mu{\~n}oz Guti{\'e}rrez}, {Hou}, {Newman}, {Palanque-Delabrouille}, {P{\^a}ris}, {Percival}, {Petitjean}, {Rich}, {Rossi}, {Schneider}, {Smith}, {Vivek}, \& {Weaver}}]{2020ApJS..250....8L}
{Lyke}, B.~W., {Higley}, A.~N., {McLane}, J.~N., {et~al.} 2020, \apjs, 250, 8

\bibitem[{Nadolny {et~al.}(2021)Nadolny, Durrer, Kunz, \& Padmanabhan}]{Nadolny_2021}
Nadolny, T., Durrer, R., Kunz, M., \& Padmanabhan, H. 2021, \jcap, 2021, 009

\bibitem[{{Oayda} {et~al.}(2024){Oayda}, {Mittal}, {Lewis}, \& {Murphy}}]{2024MNRAS.531.4545O}
{Oayda}, O.~T., {Mittal}, V., {Lewis}, G.~F., \& {Murphy}, T. 2024, \mnras, 531, 4545

\bibitem[{Park {et~al.}(2017)Park, Hyun, Noh, \& Hwang}]{10.1093/mnras/stx988}
Park, C.-G., Hyun, H., Noh, H., \& Hwang, J.-c. 2017, \mnras, 469, 1924

\bibitem[{{Planck Collaboration} {et~al.}(2016){Planck Collaboration}, {Ade, P. A. R.}, {Aghanim, N.}, {Akrami, Y.}, {Aluri, P. K.}, {Arnaud, M.}, {Ashdown, M.}, {Aumont, J.}, {Baccigalupi, C.}, {Banday, A. J.}, {Barreiro, R. B.}, {Bartolo, N.}, {Basak, S.}, {Battaner, E.}, {Benabed, K.}, {Benoît, A.}, {Benoit-Lévy, A.}, {Bernard, J.-P.}, {Bersanelli, M.}, {Bielewicz, P.}, {Bock, J. J.}, {Bonaldi, A.}, {Bonavera, L.}, {Bond, J. R.}, {Borrill, J.}, {Bouchet, F. R.}, {Boulanger, F.}, {Bucher, M.}, {Burigana, C.}, {Butler, R. C.}, {Calabrese, E.}, {Cardoso, J.-F.}, {Casaponsa, B.}, {Catalano, A.}, {Challinor, A.}, {Chamballu, A.}, {Chiang, H. C.}, {Christensen, P. R.}, {Church, S.}, {Clements, D. L.}, {Colombi, S.}, {Colombo, L. P. L.}, {Combet, C.}, {Contreras, D.}, {Couchot, F.}, {Coulais, A.}, {Crill, B. P.}, {Cruz, M.}, {Curto, A.}, {Cuttaia, F.}, {Danese, L.}, {Davies, R. D.}, {Davis, R. J.}, {de Bernardis, P.}, {de Rosa, A.}, {de Zotti, G.}, {Delabrouille, J.}, {Désert, F.-X.}, {Diego, J. M.}, {Dole,
  H.}, {Donzelli, S.}, {Doré, O.}, {Douspis, M.}, {Ducout, A.}, {Dupac, X.}, {Efstathiou, G.}, {Elsner, F.}, {Enßlin, T. A.}, {Eriksen, H. K.}, {Fantaye, Y.}, {Fergusson, J.}, {Fernandez-Cobos, R.}, {Finelli, F.}, {Forni, O.}, {Frailis, M.}, {Fraisse, A. A.}, {Franceschi, E.}, {Frejsel, A.}, {Frolov, A.}, {Galeotta, S.}, {Galli, S.}, {Ganga, K.}, {Gauthier, C.}, {Ghosh, T.}, {Giard, M.}, {Giraud-Héraud, Y.}, {Gjerløw, E.}, {González-Nuevo, J.}, {Górski, K. M.}, {Gratton, S.}, {Gregorio, A.}, {Gruppuso, A.}, {Gudmundsson, J. E.}, {Hansen, F. K.}, {Hanson, D.}, {Harrison, D. L.}, {Henrot-Versillé, S.}, {Hernández-Monteagudo, C.}, {Herranz, D.}, {Hildebrandt, S. R.}, {Hivon, E.}, {Hobson, M.}, {Holmes, W. A.}, {Hornstrup, A.}, {Hovest, W.}, {Huang, Z.}, {Huffenberger, K. M.}, {Hurier, G.}, {Jaffe, A. H.}, {Jaffe, T. R.}, {Jones, W. C.}, {Juvela, M.}, {Keihänen, E.}, {Keskitalo, R.}, {Kim, J.}, {Kisner, T. S.}, {Knoche, J.}, {Kunz, M.}, {Kurki-Suonio, H.}, {Lagache, G.}, {Lähteenmäki, A.}, {Lamarre,
  J.-M.}, {Lasenby, A.}, {Lattanzi, M.}, {Lawrence, C. R.}, {Leonardi, R.}, {Lesgourgues, J.}, {Levrier, F.}, {Liguori, M.}, {Lilje, P. B.}, {Linden-Vørnle, M.}, {Liu, H.}, {López-Caniego, M.}, {Lubin, P. M.}, {Macías-Pérez, J. F.}, {Maggio, G.}, {Maino, D.}, {Mandolesi, N.}, {Mangilli, A.}, {Marinucci, D.}, {Maris, M.}, {Martin, P. G.}, {Martínez-González, E.}, {Masi, S.}, {Matarrese, S.}, {McGehee, P.}, {Meinhold, P. R.}, {Melchiorri, A.}, {Mendes, L.}, {Mennella, A.}, {Migliaccio, M.}, {Mikkelsen, K.}, {Mitra, S.}, {Miville-Deschênes, M.-A.}, {Molinari, D.}, {Moneti, A.}, {Montier, L.}, {Morgante, G.}, {Mortlock, D.}, {Moss, A.}, {Munshi, D.}, {Murphy, J. A.}, {Naselsky, P.}, {Nati, F.}, {Natoli, P.}, {Netterfield, C. B.}, {Nørgaard-Nielsen, H. U.}, {Noviello, F.}, {Novikov, D.}, {Novikov, I.}, {Oxborrow, C. A.}, {Paci, F.}, {Pagano, L.}, {Pajot, F.}, {Pant, N.}, {Paoletti, D.}, {Pasian, F.}, {Patanchon, G.}, {Pearson, T. J.}, {Perdereau, O.}, {Perotto, L.}, {Perrotta, F.}, {Pettorino, V.},
  {Piacentini, F.}, {Piat, M.}, {Pierpaoli, E.}, {Pietrobon, D.}, {Plaszczynski, S.}, {Pointecouteau, E.}, {Polenta, G.}, {Popa, L.}, {Pratt, G. W.}, {Prézeau, G.}, {Prunet, S.}, {Puget, J.-L.}, {Rachen, J. P.}, {Rebolo, R.}, {Reinecke, M.}, {Remazeilles, M.}, {Renault, C.}, {Renzi, A.}, {Ristorcelli, I.}, {Rocha, G.}, {Rosset, C.}, {Rossetti, M.}, {Rotti, A.}, {Roudier, G.}, {Rubiño-Martín, J. A.}, {Rusholme, B.}, {Sandri, M.}, {Santos, D.}, {Savelainen, M.}, {Savini, G.}, {Scott, D.}, {Seiffert, M. D.}, {Shellard, E. P. S.}, {Souradeep, T.}, {Spencer, L. D.}, {Stolyarov, V.}, {Stompor, R.}, {Sudiwala, R.}, {Sunyaev, R.}, {Sutton, D.}, {Suur-Uski, A.-S.}, {Sygnet, J.-F.}, {Tauber, J. A.}, {Terenzi, L.}, {Toffolatti, L.}, {Tomasi, M.}, {Tristram, M.}, {Trombetti, T.}, {Tucci, M.}, {Tuovinen, J.}, {Valenziano, L.}, {Valiviita, J.}, {Van Tent, B.}, {Vielva, P.}, {Villa, F.}, {Wade, L. A.}, {Wandelt, B. D.}, {Wehus, I. K.}, {Yvon, D.}, {Zacchei, A.}, {Zibin, J. P.}, \& {Zonca, A.}}]{refId01}
{Planck Collaboration}, {Ade, P. A. R.}, {Aghanim, N.}, {et~al.} 2016, \aap, 594, A16

\bibitem[{{Planck Collaboration} {et~al.}(2020){Planck Collaboration}, {Akrami, Y.}, {Ashdown, M.}, {Aumont, J.}, {Baccigalupi, C.}, {Ballardini, M.}, {Banday, A. J.}, {Barreiro, R. B.}, {Bartolo, N.}, {Basak, S.}, {Benabed, K.}, {Bersanelli, M.}, {Bielewicz, P.}, {Bock, J. J.}, {Bond, J. R.}, {Borrill, J.}, {Bouchet, F. R.}, {Boulanger, F.}, {Bucher, M.}, {Burigana, C.}, {Butler, R. C.}, {Calabrese, E.}, {Cardoso, J.-F.}, {Casaponsa, B.}, {Chiang, H. C.}, {Colombo, L. P. L.}, {Combet, C.}, {Contreras, D.}, {Crill, B. P.}, {de Bernardis, P.}, {de Zotti, G.}, {Delabrouille, J.}, {Delouis, J.-M.}, {Di Valentino, E.}, {Diego, J. M.}, {Doré, O.}, {Douspis, M.}, {Ducout, A.}, {Dupac, X.}, {Efstathiou, G.}, {Elsner, F.}, {Enßlin, T. A.}, {Eriksen, H. K.}, {Fantaye, Y.}, {Fernandez-Cobos, R.}, {Finelli, F.}, {Frailis, M.}, {Fraisse, A. A.}, {Franceschi, E.}, {Frolov, A.}, {Galeotta, S.}, {Galli, S.}, {Ganga, K.}, {Génova-Santos, R. T.}, {Gerbino, M.}, {Ghosh, T.}, {González-Nuevo, J.}, {Górski, K. M.},
  {Gruppuso, A.}, {Gudmundsson, J. E.}, {Hamann, J.}, {Handley, W.}, {Hansen, F. K.}, {Herranz, D.}, {Hivon, E.}, {Huang, Z.}, {Jaffe, A. H.}, {Jones, W. C.}, {Keihänen, E.}, {Keskitalo, R.}, {Kiiveri, K.}, {Kim, J.}, {Krachmalnicoff, N.}, {Kunz, M.}, {Kurki-Suonio, H.}, {Lagache, G.}, {Lamarre, J.-M.}, {Lasenby, A.}, {Lattanzi, M.}, {Lawrence, C. R.}, {Le Jeune, M.}, {Levrier, F.}, {Liguori, M.}, {Lilje, P. B.}, {Lindholm, V.}, {López-Caniego, M.}, {Ma, Y.-Z.}, {Macías-Pérez, J. F.}, {Maggio, G.}, {Maino, D.}, {Mandolesi, N.}, {Mangilli, A.}, {Marcos-Caballero, A.}, {Maris, M.}, {Martin, P. G.}, {Martínez-González, E.}, {Matarrese, S.}, {Mauri, N.}, {McEwen, J. D.}, {Meinhold, P. R.}, {Mennella, A.}, {Migliaccio, M.}, {Miville-Deschênes, M.-A.}, {Molinari, D.}, {Moneti, A.}, {Montier, L.}, {Morgante, G.}, {Moss, A.}, {Natoli, P.}, {Pagano, L.}, {Paoletti, D.}, {Partridge, B.}, {Perrotta, F.}, {Pettorino, V.}, {Piacentini, F.}, {Polenta, G.}, {Puget, J.-L.}, {Rachen, J. P.}, {Reinecke, M.},
  {Remazeilles, M.}, {Renzi, A.}, {Rocha, G.}, {Rosset, C.}, {Roudier, G.}, {Rubiño-Martín, J. A.}, {Ruiz-Granados, B.}, {Salvati, L.}, {Savelainen, M.}, {Scott, D.}, {Shellard, E. P. S.}, {Sirignano, C.}, {Sunyaev, R.}, {Suur-Uski, A.-S.}, {Tauber, J. A.}, {Tavagnacco, D.}, {Tenti, M.}, {Toffolatti, L.}, {Tomasi, M.}, {Trombetti, T.}, {Valenziano, L.}, {Valiviita, J.}, {Van Tent, B.}, {Vielva, P.}, {Villa, F.}, {Vittorio, N.}, {Wandelt, B. D.}, {Wehus, I. K.}, {Zacchei, A.}, {Zibin, J. P.}, \& {Zonca, A.}}]{refId0}
{Planck Collaboration}, {Akrami, Y.}, {Ashdown, M.}, {et~al.} 2020, \aap, 641, A7

\bibitem[{RALSTON \& JAIN(2004)}]{RALSTON_2004}
RALSTON, J.~P. \& JAIN, P. 2004, International Journal of Modern Physics D, 13, 1857–1877

\bibitem[{Rameez {et~al.}(2018)Rameez, Mohayaee, Sarkar, \& Colin}]{Rameez_2018}
Rameez, M., Mohayaee, R., Sarkar, S., \& Colin, J. 2018, \mnras, 477, 1772–1781

\bibitem[{Ross {et~al.}(2020)Ross, Bautista, Tojeiro, Alam, Bailey, Burtin, Comparat, Dawson, de Mattia, du Mas des Bourboux, Gil-Marín, Hou, Kong, Lyke, Mohammad, Moustakas, Mueller, Myers, Percival, Raichoor, Rezaie, Seo, Smith, Tinker, Zarrouk, Zhao, Zhao, Bizyaev, Brinkmann, Brownstein, Rosell, Chabanier, Choi, Chuang, Cruz-Gonzalez, de la Macorra, de la Torre, Escoffier, Fromenteau, Higley, Jullo, Kneib, McLane, Muñoz-Gutiérrez, Neveux, Newman, Nitschelm, Palanque-Delabrouille, Paviot, Pullen, Rossi, Ruhlmann-Kleider, Schneider, Magaña, Vivek, \& Zhang}]{10.1093/mnras/staa2416}
Ross, A.~J., Bautista, J., Tojeiro, R., {et~al.} 2020, \mnras, 498, 2354

\bibitem[{Rubart {et~al.}(2014)Rubart, Bacon, \& Schwarz}]{Rubart_2014}
Rubart, M., Bacon, D., \& Schwarz, D.~J. 2014, \aap, 565, A111

\bibitem[{Rubart \& Schwarz(2013)}]{Rubart_2013}
Rubart, M. \& Schwarz, D.~J. 2013, \aap, 555, A117

\bibitem[{Sarkar {et~al.}(2018)Sarkar, Pandey, \& Khatri}]{Sarkar_2018}
Sarkar, S., Pandey, B., \& Khatri, R. 2018, \mnras, 483, 2453–2464

\bibitem[{Secrest {et~al.}(2021)Secrest, Hausegger, Rameez, Mohayaee, Sarkar, \& Colin}]{Secrest_2021}
Secrest, N.~J., Hausegger, S.~v., Rameez, M., {et~al.} 2021, \apjl, 908, L51

\bibitem[{Secrest {et~al.}(2022)Secrest, von Hausegger, Rameez, Mohayaee, \& Sarkar}]{Secrest_2022}
Secrest, N.~J., von Hausegger, S., Rameez, M., Mohayaee, R., \& Sarkar, S. 2022, \apjl, 937, L31

\bibitem[{Singal(2011)}]{Singal_2011}
Singal, A.~K. 2011, \apj, 742, L23

\bibitem[{Singal(2022)}]{Singal_2022}
Singal, A.~K. 2022, \mnras, 515, 5969–5980

\bibitem[{Sorrenti {et~al.}(2023)Sorrenti, Durrer, \& Kunz}]{sorrenti2023dipolepantheonsh0esdata}
Sorrenti, F., Durrer, R., \& Kunz, M. 2023, The Dipole of the Pantheon+SH0ES Data

\bibitem[{Tiwari \& Nusser(2016)}]{Tiwari_2016}
Tiwari, P. \& Nusser, A. 2016, \jcap, 2016, 062–062

\bibitem[{Wu {et~al.}(1999)Wu, Lahav, \& Rees}]{Wu_1999}
Wu, K. K.~S., Lahav, O., \& Rees, M.~J. 1999, Nature, 397, 225–230

\bibitem[{Yang {et~al.}(2013)Yang, Wang, \& Chu}]{Yang_2013}
Yang, X., Wang, F.~Y., \& Chu, Z. 2013, \mnras, 437, 1840–1846

\end{thebibliography}

\appendix
\section{Methodological validation and additional analyses}

This appendix provides supporting details for the dipole analysis presented in the main text, focusing on methodological consistency and the exploration of potential systematics.

As a primary validation step of our entire analysis pipeline---including data processing, dipole estimation, and uncertainty quantification---we applied it to the eBOSS DR16 QSO sample. This is the same dataset used in the benchmark study of \citep{da_Silveira_Ferreira_2024}. 

We followed the identical procedure outlined in Section~\ref{sec:method} of the main text. The sample selection, coordinate transformations, and likelihood analysis for the dipole components were performed without modification.

\begin{figure}[!htbp]
    \centering
    \includegraphics[width=\columnwidth]{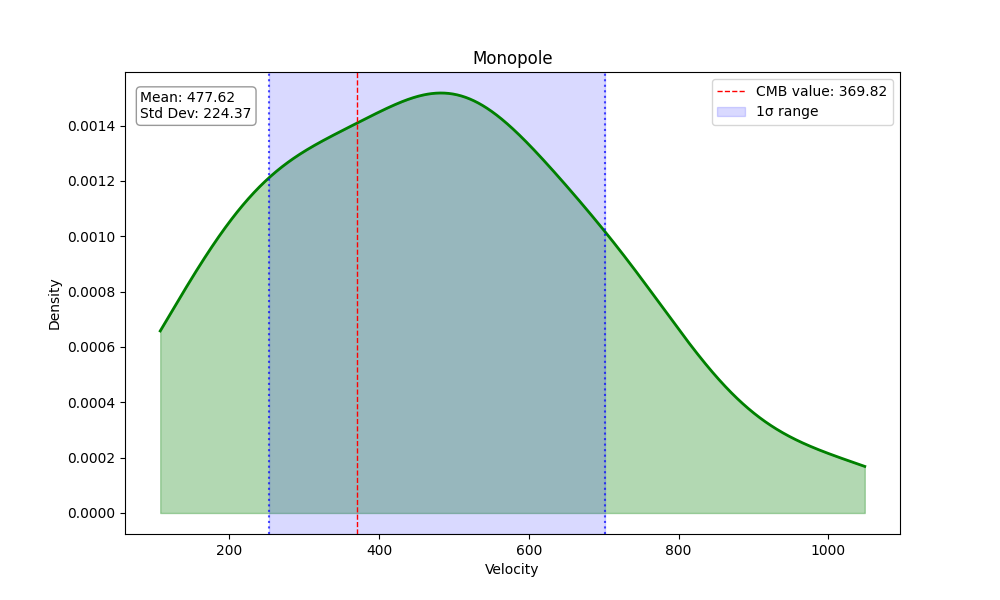}
    \includegraphics[width=\columnwidth]{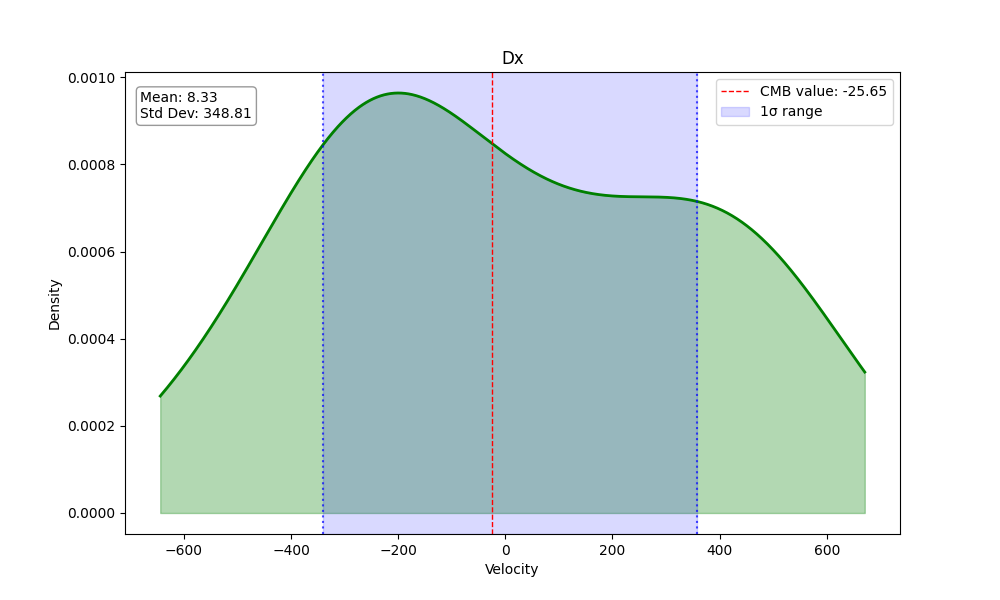}
    \includegraphics[width=\columnwidth]{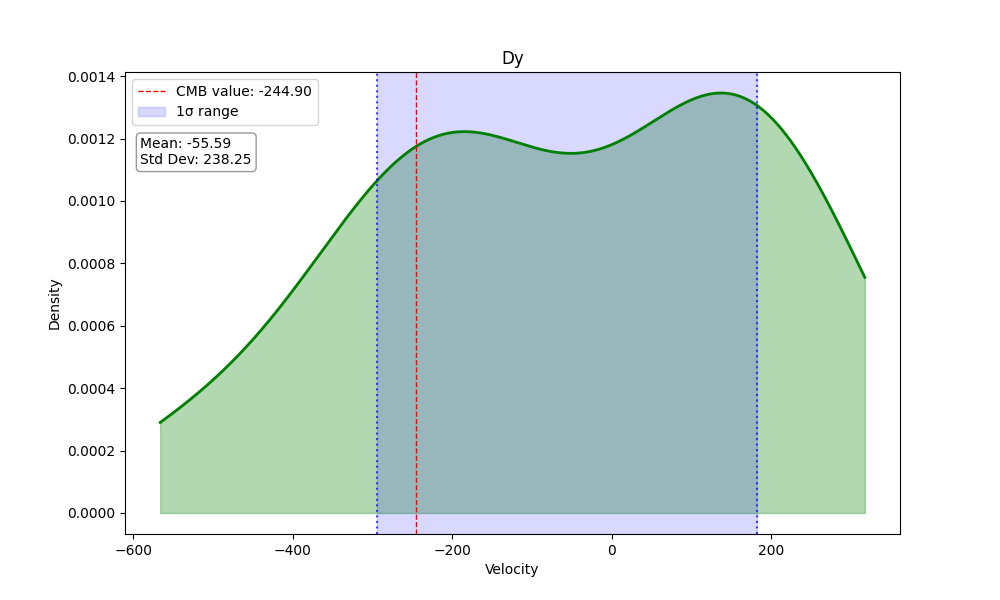}
    \includegraphics[width=\columnwidth]{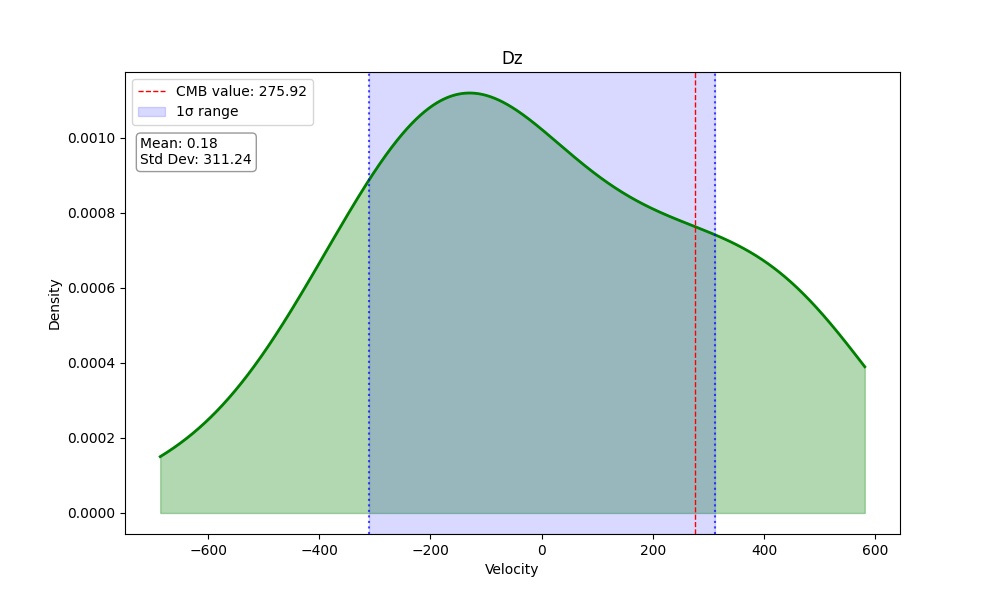}
    \caption{Empirical distribution plots of the eBOSS DR16 redshift dipole in its Cartesian components (m, $c\Delta_x$, $c\Delta_y$, $c\Delta_z$). The red curve represents the dipole expected from the CMB dipole. The x-axis represents velocity (in km/s), the y-axis represents distribution density, and the shaded regions denote the $1\sigma$ confidence intervals. Notably, all components   fully encompass the CMB dipole value at the $1\sigma$ ranges.}
    \label{appfig1}
\end{figure}

The resulting probability distributions for the monopole and the three Cartesian dipole components are shown in Fig.~\ref{appfig1} in the main text. The key outcome is that the expected value from the CMB kinematic dipole (red line) falls entirely within the $1\sigma$ posterior credible interval for every single component.

This successful replication serves two crucial purposes:
1. Methodological Robustness: It confirms that our implementation of the \citep{da_Silveira_Ferreira_2024} method is correct and yields consistent results on the same data.
2. Baseline Establishment: It provides a clear benchmark for a “consistent” result within the $\Lambda$CDM framework. The contrast between this clean consistency (eBOSS DR16) and the component-level tension found in DESI DR1 (Fig.~\ref{fig:3}) sharpens the discussion on the origin of the latter.

\end{document}